\documentclass[sigconf,10pt]{acmart}
\AtBeginDocument{%
  }

\copyrightyear{2026}
\acmYear{2026}
\setcopyright{cc}
\setcctype{by}
\acmConference[MobiCom '26]{The 32nd Annual International Conference on Mobile Computing and Networking}{October 26--30, 2026}{Austin, TX, USA}
\acmBooktitle{The 32nd Annual International Conference on Mobile Computing and Networking (MobiCom '26), October 26--30, 2026, Austin, TX, USA}
\acmPrice{}
\acmDOI{10.1145/3795866.3796700}
\acmISBN{979-8-4007-2505-0/2026/10}



\usepackage{amsmath}
\usepackage{makecell}
\usepackage{enumerate}
\usepackage{changepage}
\usepackage{dsfont}
\usepackage{pifont}
\usepackage{gensymb}
\usepackage{charter, epsfig, endnotes, verbatim, color}
\usepackage{xcolor}
\usepackage{multirow, soul, tabularx}
\usepackage{textcomp}
\usepackage{wrapfig}
\usepackage{algorithm}
\usepackage{algorithmic}
\usepackage[inline]{enumitem}
\usepackage{array,etoolbox}
\usepackage{float}
\usepackage{balance}
\usepackage{tablefootnote}
\usepackage[font=small,labelfont=bf,tableposition=top]{caption}
\usepackage{subfig}
\usepackage{colortbl}
\usepackage{pifont}
\usepackage{booktabs}
\usepackage{siunitx}
\usepackage{tikz}

\usepackage{mdframed}
\usepackage[most]{tcolorbox}

\definecolor{Gray}{gray}{0.8}
\definecolor{lightgreen}{HTML}{DDE6D7}
\definecolor{lightblue}{HTML}{D6DDEA}
\definecolor{lightred}{HTML}{e9967a}

\newcommand{\name}{\emph{Press2Muscle}}

\def\task #1#2 {$\small{{#1}}2\small{{#2}}$}
\def\taskbf #1#2 {$\small{\textbf{#1}} 2 \small{\textbf{#2}}$}

\newcommand{\lt}{\ensuremath <}
\newcommand{\gt}{\ensuremath >}

\newcommand{\rev}{\textcolor{black}}

\newcommand{\hao}{\textcolor{black}}
\newcounter{boxctr}

\title{Exploring the Feasibility of Full-Body Muscle Activation Sensing with Insole Pressure Sensors}

\author{Hao Zhou and Mahanth Gowda}
\affiliation{
  \institution{The Pennsylvania State University} 
  \country{}
}

\renewcommand{\shortauthors}{H. Zhou, M. Gowda}

\begin{document}

\renewcommand{\thetable}{S\arabic{table}}
\renewcommand{\thefigure}{S\arabic{figure}}

\renewcommand{\thetable}{\arabic{table}}
\renewcommand{\thefigure}{\arabic{figure}}

\setcounter{figure}{0}
\setcounter{page}{1}

\begin{abstract}
Muscle activation initiates contractions that drive human movement, and understanding it provides valuable insights for injury prevention and rehabilitation. Yet, sensing muscle activation is barely explored in the rapidly growing mobile health market. Traditional methods for muscle activation sensing rely on specialized electrodes, such as surface electromyography, making them impractical, especially for long-term usage. 
In this paper, we introduce {\name}, the first system to unobtrusively infer muscle activation using insole pressure sensors. The key idea is to analyze foot pressure changes resulting from full-body muscle activation that drives movements. To handle variations in pressure signals due to differences in users' gait, weight, and movement styles, we propose a data-driven approach to dynamically adjust reliance on different foot regions and incorporate easily accessible biographical data %
to enhance {\name}’s generalization to unseen users. We conducted an extensive study with 30 users. Under a leave-one-user-out setting, {\name} achieves a root mean square error of 0.025, marking a 19\% improvement over a video-based counterpart. %
A robustness study validates {\name}’s ability to generalize across user demographics, footwear types, and walking surfaces. \rev{Additionally, we showcase muscle imbalance detection and muscle activation estimation under free-living settings with {\name}, confirming the feasibility of muscle activation sensing using insole pressure sensors in real-world settings.}

\end{abstract}

\maketitle

\begin{figure}[hbt]

    \centering
\includegraphics[width=1\columnwidth, scale=1]{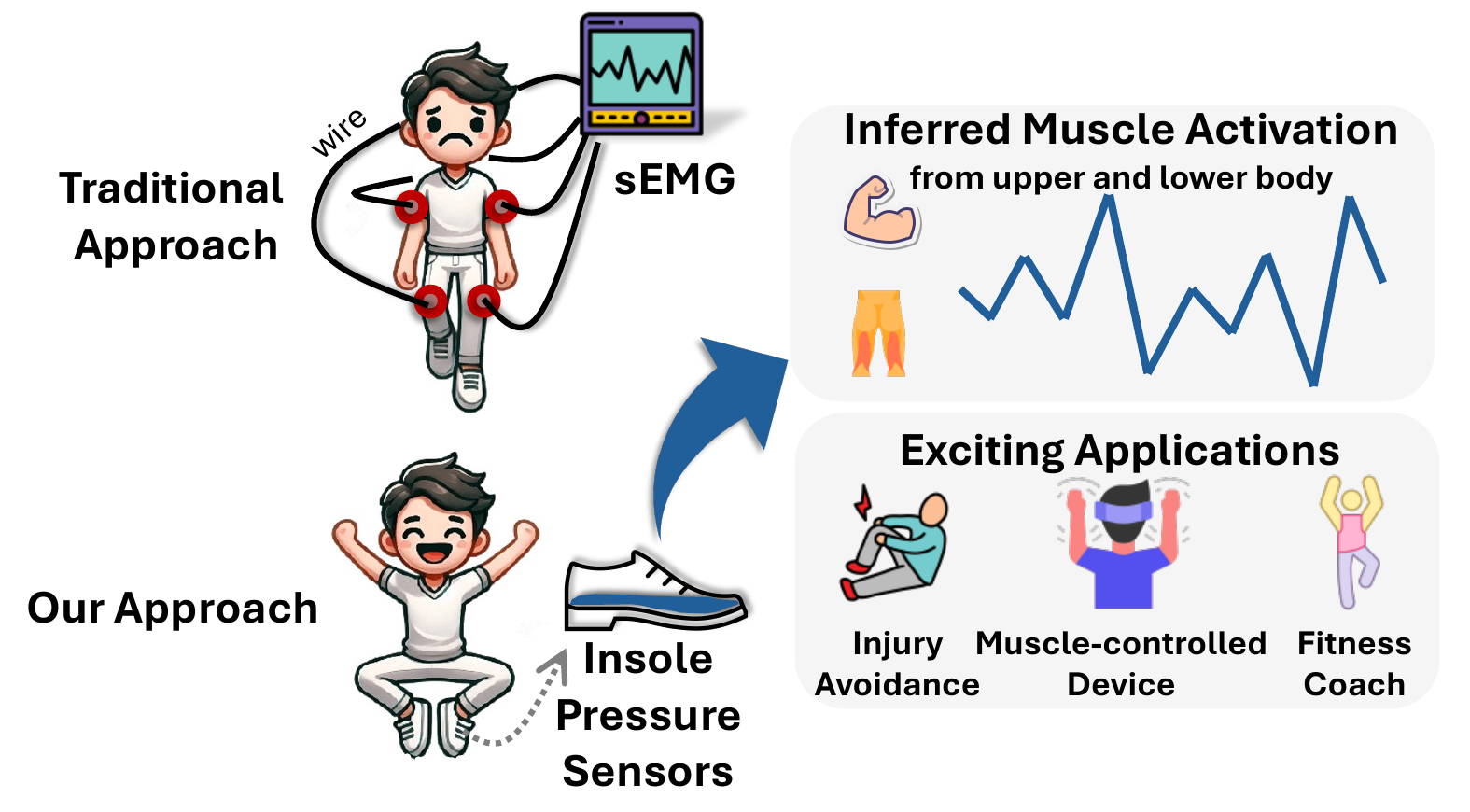}
    \caption{\rev{In contrast to traditional methods, {\name} provides an unobtrusive and scalable solution for accurate full-body muscle activation monitoring with pressure insoles, unlocking broad applications in health and fitness.}}
    \label{fig:motivation}
\end{figure}

\section{Introduction}\label{sec:intro}

Imagine Emma, a running enthusiast, training for her first half-marathon. Despite her efforts, knee pain started, which she dismissed as muscle fatigue until a sharp pain during her practice forced her to stop. A physiotherapist later found the cause: her quads were overworking while other muscles failed to coordinate. If monitoring muscle activity across the whole body were accessible in daily life, unnecessary injuries could be avoided, just as numerous works
that track physiological status (e.g., heart rate and stress) to alert users with potential health issues~\cite{health_hr1,zeng2019farsense, health_bp1, health_emotion1,health_res3, zhou2025rethinking, lee2025himae}. 
\rev{As the research community and major tech companies continue advancing health technologies, integrating muscle activation monitoring could revolutionize how we monitor physical well-being, opening a wide range of applications, including injury recovery~\cite{rehab1,baoge2012treatment,tidball2005inflammatory}, personalized coaching for optimizing movement patterns, particularly in virtual reality environments~\cite{gait1,wu2025muscle, zhou2022learning}.} \\[-0.7em]

\noindent \textbf{Prior Work}. Traditional methods of measuring muscle activity primarily rely on specialized equipment such as Surface Electromyography (sEMG)~\cite{emg1, emg3, emg4}. While they are the most accurate approaches to measure muscle activation, the attachments of specialized electrodes make them expensive or cause skin irritation and discomfort~\cite{emg2}. While videos inherently do not sense any muscular activity, video-based approaches~\cite{mit, mia} provide an alternative by building connections between muscle activity and external body movements. Yet, videos are not ubiquitous, often limited to lighting conditions, occlusions, and privacy concerns~\cite{zhao2018through,zhang2021widar3,liu2021wr,cao2024finger}. \\[-0.7em]

\noindent \textbf{New Opportunity}. \rev{Muscle contractions drive human motions, and every overt or subtle motion performed generally leads to dynamic shifts of the human body~\cite{mia, cruz2025wrist, hong2020effects, ustinova2013postural}. This weight re-distribution can be directly sensed through the feet, as the forces exerted on the ground reflect biomechanical adjustments driven by muscle activation patterns throughout the body~\cite{ustinova2013postural,hicks2015my, bostrom2018contribution, hong2020effects}. This intrinsic connection (Sec.~\ref{sec:just}) between muscle activation and foot pressure dynamics 
presents us with a new opportunity. Leveraging this opportunity, we propose {\name}, an unobtrusive and scalable solution for full-body muscle activation (eight muscles from the upper and lower body with details in Sec.~\ref{sec:setup}) sensing with insole pressure sensors (Fig.~\ref{fig:motivation}).} \\[-0.7em]

\noindent\textbf{Challenges}. We face multiple challenges: \\[-0.7em]

\noindent $\bullet$ Foot pressure patterns vary widely across different users due to physiological and biomechanical differences~\cite{gait5,gait6}. Factors such as weight, height, gait style, and muscle strength all influence pressure distribution. For instance, a lighter individual exerts less force overall, whereas a heavier user generates consistently higher pressure values across all movements. Additionally, we observe that some users apply greater force on their heels when in motion, while others distribute pressure more evenly across the feet. Such user-specific variations make it challenging to develop a universal model that performs well across all individuals without per-user adaptation training. \\[-0.7em]

\noindent$\bullet$ Even within the same individual, foot pressure patterns fluctuate across different conditions due to factors such as footwear, walking surfaces, and natural variations in movement execution~\cite{bruijn2018control,kowalsky2021human}. We observe that different shoes may alter how pressure is distributed across feet; softer surfaces (e.g., carpet) attenuate pressure responses compared to harder surfaces (e.g., concrete); when a user alternately pushes his legs, the pressure responses from each repetition are not identical due to stability adjustments, etc. Such variability presents a unique challenge, as {\name} needs to be reliable and accurate across diverse real-world conditions.  \\[-0.7em]

\noindent \textbf{Solutions}. We propose a novel learning network for foot pressure data and incorporate easily accessible biographical information, enabling effective user adaptation with zero intervention. 
Furthermore, we propose pressure data augmentations to enhance {\name} adaptability across diverse conditions.  \\[-0.7em]

\noindent $\bullet$ We propose \emph{Region Importance Learning} that allows {\name} to dynamically adjust its reliance on different foot regions by developing a learnable mask. This mask allows {\name} to focus on informative regions from highly dynamic pressure data while reducing sensitivity to irrelevant signals (e.g., consistently exhibit zero values due to limited ground contact in Sec.~\ref{sec:prelimnary}).
Additionally, to deal with the challenge of varying foot pressure patterns across individuals due to differences in weight, height, and gait style, we integrate readily available biographical information (e.g., weight, height, and gender) to dynamically scale the pressure data, enabling effective user adaptation with nearly-free cost. In combination, {\name} effectively adapts across users with diverse heights, weights, etc. \\[-0.7em]

\noindent$\bullet$ We introduce simple yet effective augmentation techniques driven by our observations of pressure data. 1) \emph{Pressure Scaling}: We apply varying scaling factors to foot regions to simulate differences in pressure sensor readings due to factors like footwear, walking surfaces, and slight movement inconsistencies. 2) \emph{Temporal Shifting}: We shift pressure data forward or backward in time while preserving the overall sequence structure. These augmentation techniques ensure {\name} learns from diverse pressure distributions without overfitting to a specific user pattern or an environmental condition, and make {\name} robust in real-world conditions. \\[-0.7em]

\noindent\textbf{Overview.} To our knowledge, {\name} is the first to explore muscle activation sensing using insole pressure sensors, offering an unobtrusive and scalable solution with broad applications in mobile health. \rev{To validate {\name}, we conducted an extensive study with 30 users. Under a leave-one-user-out setting, {\name} achieves a root mean square error of 0.025 when estimating the activity of eight frequently-used muscle groups across the body, demonstrating generalization across users with diverse demographics, including age, muscle composition, foot type, and gender. Additionally, we conducted robustness studies to assess {\name} under various real-world conditions, including motion types and walking surfaces. We also showcased muscle imbalance detection and estimation of muscle activation under in-situ settings, highlighting {\name}'s practicability.} We summarize our contributions as follows.  \\[-0.7em]
\begin{itemize}[leftmargin=*]
    \item \rev{{\name} is the first work that explores the feasibility of muscle activation sensing using insole pressure sensors, leveraging the intrinsic connection between muscle activation and foot pressure.}
    \item Through a deep understanding of foot pressure, we propose \emph{region importance learning} to focus on informative regions, alongside \emph{readily available biographical data} and \emph{pressure-specific augmentations} for better generalization across users and diverse conditions.
    \item \rev{We demonstrate {\name}'s effectiveness across 30 users with diverse demographics and practicability via muscle imbalance detection and in-situ activities. } 
\end{itemize}

\section{Opportunity, Theoretical Analysis, Vision}\label{sec:bg}

In this section, we present the opportunity that motivates {\name}, a theoretical analysis, and a list of potential applications with {\name}.

\subsection{Opportunities}
While muscle activation is central to human movement and health in daily life, it has received limited attention from academia and industry. We propose leveraging foot pressure sensors (Fig.~\ref{fig:platform}a) as a novel approach to capture muscle activity seamlessly. These sensors can be unobtrusively embedded in insoles or footwear (Fig.~\ref{fig:platform}c), ensuring ubiquity and comfort. They also preserve privacy by avoiding video recordings, while remaining cost-effective and scalable for widespread use in health monitoring, fitness analysis, and beyond.

\subsection{A Theoretical Analysis }\label{sec:just}

When a muscle $i$ is activated at level $a_i$, it produces a muscle force~\cite{zajac1989muscle}, which in turn generates joint torque $\tau_j$ at joint $j$. 
Because the human body is an interconnected musculoskeletal system, torques generated at individual joints propagate through the kinetic chain, producing coordinated segmental motions that alter limb configurations and shift the body’s center of mass (CoM)~\cite{ellenbecker2001closed,stapley1999does}. This shift can be quantitatively described by the CoM equation,
$\vec{x}_{\text{CoM}} = \frac{1}{M} \sum_i m_i \vec{x}_i$,
where $m_i$ and $\vec{x}_i$ denote the mass and position of segment $i$, and $M$ is the total body mass. 
Changes in the CoM directly influence balance and stability, prompting compensatory adjustments in the ground reaction force (GRF) and, consequently, the center of pressure (CoP) under the feet. 
In the anterior--posterior direction, the CoP location is given by
$\text{CoP}_x = \frac{-M h \ddot{x}_{\text{CoM}}}{F_z}$,
where $F_z$ is the vertical component of the GRF, $h$ is the height of the CoM above the ground, and $\ddot{x}_{\text{CoM}}$ denotes the horizontal acceleration of the CoM.
For clarity, we present the formulation in one direction, but the same relationship extends to all spatial dimensions. \hao{Importantly, CoP variations are not solely driven by lower-limb motion. Due to whole-body biomechanical coupling, upper-limb muscle activations can also induce measurable CoP shifts, even in the absence of overt lower-body joint movement.
Upper-body contractions generate internal torques that propagate through the kinetic chain, producing accelerations of the trunk and upper-body segments.
These accelerations modulate CoM dynamics and redistribute ground reaction forces at the feet, resulting in compensatory CoP displacement. From a control perspective, CoP therefore reflects not only static posture but also dynamic balance regulation, whereby the neuromuscular system continuously adjusts plantar pressure to stabilize CoM motion regardless of whether activations originate from the upper or lower body.} Such effects have been empirically observed both when arm motion is intentionally restrained~\cite{cruz2025wrist, bostrom2018contribution, hong2020effects} and during tasks involving controlled upper-body force production, such as squeezing a hand grip with varying force levels while standing~\cite{ustinova2013postural}.
Together, these observations form a unified biomechanical chain spanning muscle activations across the entire body\footnote{This analysis focuses on muscle activations without heavy external tools; tool-assisted movements are left for future work.}:

\begin{center}
\begin{tikzpicture}
\node (eq) at (0,0) {
  $\underbrace{\mathbf{a}}_{\substack{\text{Muscle} \\\\ \text{Activation}}}
  \rightarrow 
  \boldsymbol{\tau}
  \rightarrow 
  \vec{x}_{\text{CoM}}, \ddot{x}_{\text{CoM}}
  \rightarrow 
  \text{CoP}
  \rightarrow 
  \underbrace{\mathbf{P}}_{\substack{\text{Foot Pressure} \\\\ \text{Pattern}}}$
};
\draw[->, thick, bend right=27]
  (-1.7,0.2) to node[below, yshift=1pt, font=\small]{Subtle motions} (0.9,0.2);
\end{tikzpicture}
\end{center}
\rev{
This formulation establishes the relationship $\mathbf{P}=f(\mathbf{a})$, enabling pressure insoles to estimate full-body muscle activation by capturing both lower- and upper-body contributions through whole-body balance dynamics.
}

\subsection{Potential Applications}\label{sec:app}
We envision potential applications with {\name}. While each application would bring unique challenges, this paper explores the feasibility of sensing muscle activation using insole pressure sensors and leaves the exploration of these exciting applications for the future. \\[-0.7em]

\noindent$\blacksquare$ \textbf{Injury Prevention and Recovery}: Monitoring muscle activation helps detect imbalances, overuse, or underuse of specific muscle groups, reducing the risk of strains and repetitive stress injuries. {\name} could detect muscle imbalance to provide objective feedback on recovery progress, ensuring patients properly engage the right muscles after an injury.  \\[-0.7em]

\noindent$\blacksquare$ \rev{\textbf{Human-Computer Interaction and Assistive Technology}: While arm muscle has been sensed for gesture-based control in VR/AR, we believe {\name}'s unique sensing location could offer more opportunities beyond the arm muscle for VR/AR or accessibility applications, e.g., improving individuals with disabilities' quality of life by enabling muscle-controlled assistive devices.} \\[-0.7em]

\noindent$\blacksquare$ \rev{\textbf{Personal Fitness Coach}: Tracking muscle activation allows fitness lovers to fine-tune movement efficiency, ensuring optimal muscle recruitment for strength, endurance, and agility. {\name} could help in identifying underactive muscles, enabling targeted training strategies for improved performance, and providing insights without the discomfort of wearing burden sensors.}

\begin{figure}[hbt]
    \centering
\includegraphics[width=1\columnwidth, scale=1]{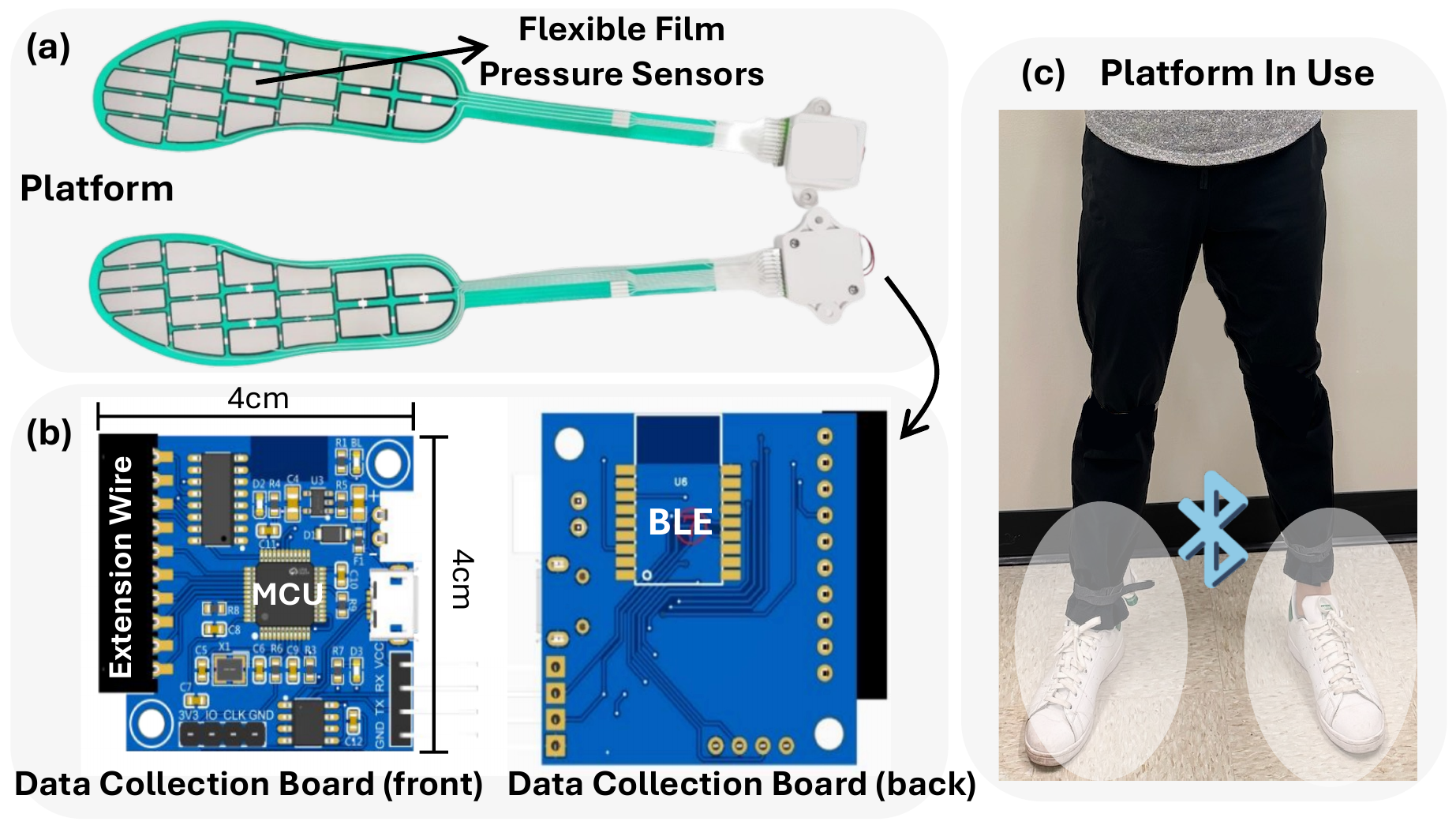}
\vspace{-0.3in}
    \caption{(a) The platform consists of multiple flexible film pressure sensors from two feet (b) with compact data collection boards that support wireless data streaming. (c) The platform can be seamlessly inserted into shoes with negligible impact on users' appearance and movements.}
    \label{fig:platform}
    \vspace{-0.1in}
\end{figure}
\section{Platform and Preliminary Study}
In this section, we first introduce the pressure sensor platform used in this project, followed by a preliminary study that suggests muscle activation induces various foot pressure responses. Finally, we share the observations that inspire the design of {\name}.

\subsection{Platform}\label{sec:platform}
The platform used in our study (Fig.~\ref{fig:platform}a) consists of 36 film pressure sensors (each foot has 18 channels). Each of the sensors operates as a pressure-sensitive variable resistor~\cite{platform1}. Thus, when pressure is applied (e.g., when a user is standing on the sensors), the sensor’s resistance decreases as the pressure increases (the relationship between resistance and pressure was pre-determined~\cite{PressInPose}). To collect pressure data, the platform uses a matrix-scanning technique, similar to a keyboard, where rows are activated sequentially, and columns are read in parallel, allowing all 18 sensors to be measured using only a few wires (Fig.~\ref{fig:platform}b). An Analog-to-Digital Converter (ADC) converts resistance changes into pressure data, which will be streamed to a laptop or smartphone via BLE (Bluetooth Low Energy), sampling at 20~Hz. The sensors priced at $\approx$80 US dollars, have a weight of $\lt$80 grams (negligible compared to the weight of standard shoes) and a thickness of $\lt$ 0.3~mm, and each of them supports a load capacity of up to 70~kg. To power the platform, a 600~mAh LiPo (Lithium Polymer) battery is used, enabling 24-hour continuous data collection and streaming. Although we did not contribute to the platform's manufacturing\footnote{The flexible film pressure sensors used in this study were purchased at \url{https://tinyurl.com/y49pean6}.}, {\name} is the first to leverage it for sensing muscle activation, which has numerous applications in health and fitness.

\begin{figure}[hbt]
    \centering
    \includegraphics[width=1\columnwidth, scale=1]{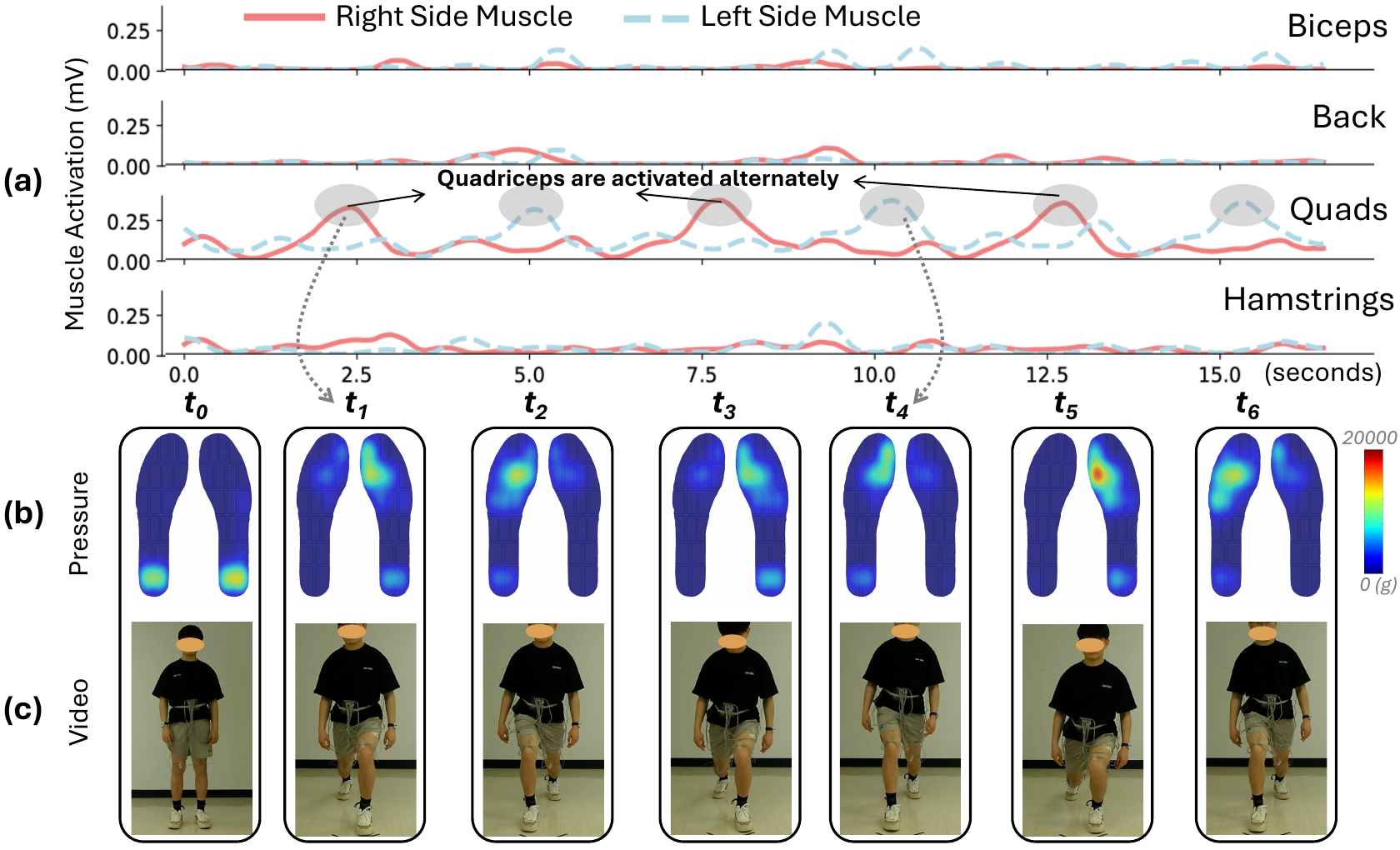}
    \vspace{-0.2in}
    \caption{\rev{An example of a user alternatively pushing his legs forward. (a) The captured activation of the quadriceps aligns with the role of knee extension, which is essential for pushing the leg forward (other muscle groups also get activated due to maintaining balance). (b) The pressure data present a coordinated pattern. 
    (c) A video is recorded to justify the movements. Note that the heatmaps are generated from 36 channels of pressure data after Gaussian interpolation for visualization purposes.}}
    \label{fig:preliminary}
\end{figure}

\subsection{Preliminary Study}\label{sec:prelimnary}

To explore whether changes in muscle activation can be reflected in foot pressure, we conducted a preliminary study where a user wore the platform while performing specific motions. To obtain ground truth muscle activation, we placed sEMG electrodes on key muscles, including the left/right biceps and hamstrings (details in Sec.~\ref{sec:setup}). We highlight an example where a user alternately pushes his legs forward. As shown in Fig.~\ref{fig:preliminary}a, the left and right quadriceps (Quads) exhibit a distinct activation pattern corresponding to the user's movements (Fig.~\ref{fig:preliminary}c), aligning with their role in knee extension, which is essential for leg propulsion. Simultaneously, pressure data (Fig.~\ref{fig:preliminary}b) reflect a coordinated pattern as weight shifts occur with knee bending. Below, we present two more interesting observations from this example. \\[-0.7em]

\noindent$\blacksquare$ \emph{Some regions of the pressure data receive little to no pressure.} This is due to variations in foot shape, individual gait patterns, etc. While this is expected, learning from inconsistent data can lead to meaningless results as these inactive regions are not always the same across users or movements (e.g., pressure data spread more to the outer foot area at $t_3$ than $t_1$). This motivates our design of region importance learning (Sec.~\ref{sec:channel}), which dynamically adjusts the importance of pressure channels (Fig.~\ref{fig:platform}a) to filter out irrelevant or inactive regions instead of treating them equally for meaningful variations. \\[-0.7em]

\noindent$\blacksquare$ \emph{Even for the same movement, the magnitude of pressure data varies.} This occurs naturally as users may exert different forces while performing the same activity (e.g., pushing harder at time $t_{5}$ than at time $t_{3}$, and more regions got activated on the left foot at $t_{6}$ than at $t_2$). Motivated by this observation, we introduce a simple yet meaningful augmentation to scale pressure data (Sec.~\ref{sec:aug}) to prevent remembering specific force magnitudes by augmenting our data to a reasonable range of variations for better generalization. \\[-0.7em]

\noindent In short, while these variations pose challenges, they also provide clear evidence that pressure data are influenced by muscle activation. This presents a great opportunity to sense muscle activation unobtrusively. Motivated by these observations, we next introduce {\name}’s design, detailing how it learns the relationship between muscle activation and pressure data.

\section{Learning in {\name}}\label{sec:model}
\begin{figure}
    \centering
    \includegraphics[width=1\columnwidth, scale=1]{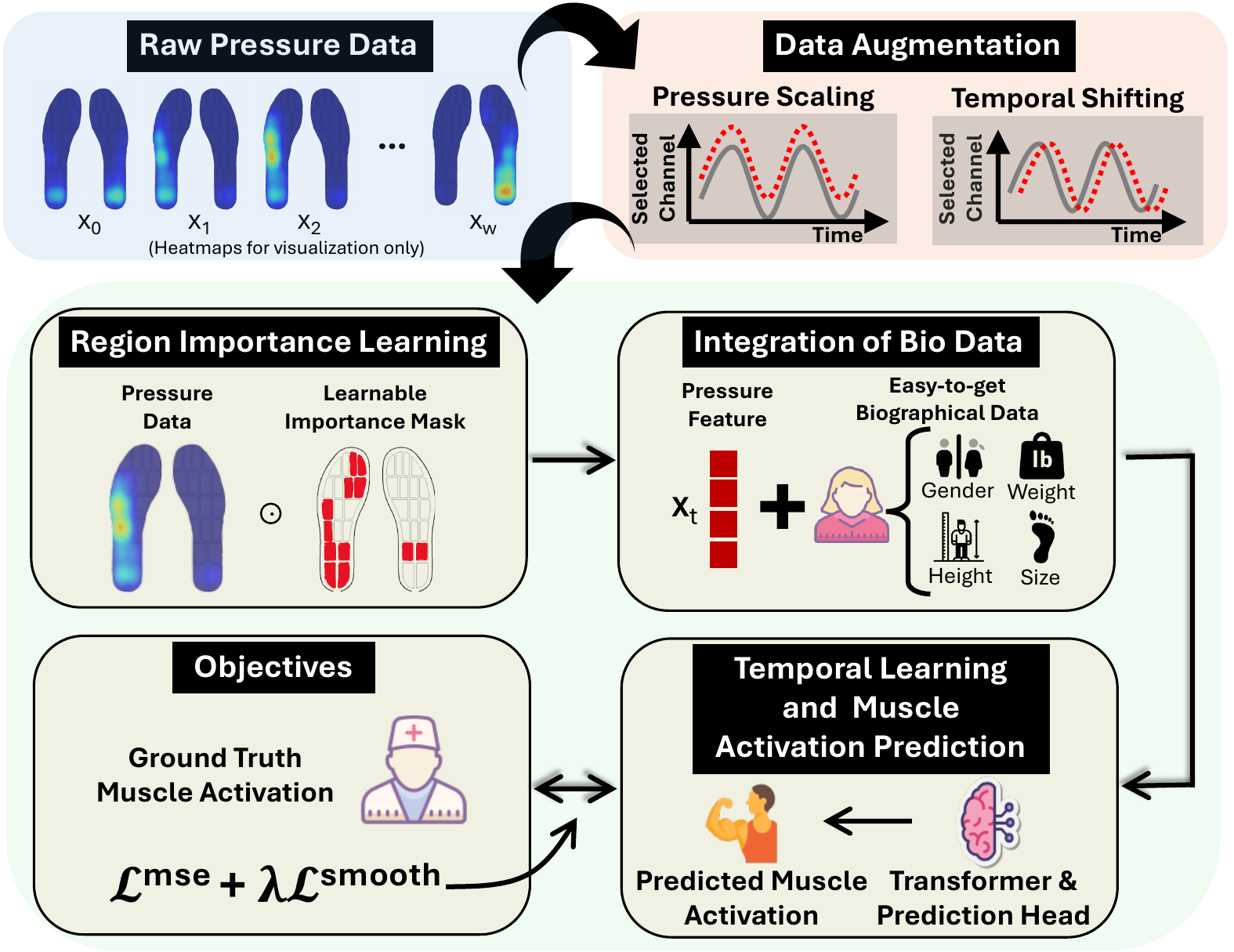}
    \vspace{-0.3in}
    \caption{{\name} Overview.}
    \label{fig:modelfig}
    \vspace{-0.1in}
\end{figure}

As depicted in Fig.~\ref{fig:modelfig}, {\name} aims to predict complex muscle activation from foot pressure data. While {\name} leverages the
intrinsic connection between foot
pressure dynamics and muscle activation (Sec.~\ref{sec:just}). Variations from users and environmental factors, such as walking surfaces, result in difficulties for a unified model. To avoid user-specific or environment-specific training, {\name} first enhances dataset diversity by proposed scaling and shifting techniques (Sec.~\ref{sec:aug}). We then introduce region importance learning to effectively capture pressure dynamics relevant to muscle activation (Sec.~\ref{sec:channel}). Furthermore, {\name} integrates readily available biographical information to improve the model’s awareness of physiological differences in users (Sec.~\ref{sec:bio}). We also learn the temporal relationship of pressure data for context muscle activation learning (Sec.~\ref{sec:temporal}), and lastly {\name}'s learning network is optimized to ensure the prediction of meaningful and smooth muscle activation (Sec.~\ref{sec:loss}). We start this section by formulating the task and sensor data preprocessing, followed by the details of these proposed techniques.

\subsection{Task Formulation and Preprocessing}\label{sec:task}

\textbf{Task Definition}: \rev{Given that foot pressure $\mathbf{P}$ is ultimately a function of upstream muscle activation $\mathbf{a}$ through the biomechanical chain (Sec.~\ref{sec:just}): $\mathbf{P} = f(\mathbf{a})$, the goal of {\name} is to solve the inverse problem of estimating the underlying muscle activation that caused the observed foot pressure distribution: 
$ \hat{\mathbf{a}} = f^{-1}(\mathbf{P})$. We model this inverse function via $f_{\theta}$, parameterized by $\theta$. } \\[-0.7em]

\noindent\textbf{Preparing Inputs}: \rev{We preprocess the raw pressure signals $\mathbf{P} \in \mathbb{R}^{T \times 2 \times 18}$, where $T$ denotes the time axis, and each sample includes measurements from both feet (18 channels per foot). The signals are segmented into fixed-length windows ($\mathbf{x}_{1:W}$), and each window is normalized to the range $[-1, 1]$ to ensure consistency across subjects and sessions. For the paired ground-truth signals $\mathbf{a} \in \mathbb{R}^{T \times 8}$ where $8$ denotes eight different muscle groups (Sec.~\ref{sec:setup}), we aligned them with the pressure signals using synchronized timestamps, 
then segmented them ($\mathbf{y}_{1:W}$) to ensure temporal consistency with the pressure input window. 
Overall, given a pressure window over $W$ time steps, our model predicts how the muscles activate over this time frame. Formally, we have
$
    f_{\theta}(\mathbf{x}_{1:W}) = \mathbf{\hat{y}} \; \; \in \mathbb{R}^{8 \times W}.
$}

\subsection{Learning Network}
As demonstrated in Sec.~\ref{sec:prelimnary}, our preliminary study revealed that certain foot pressure regions, such as the foot arch and toes, often exhibit minimal or no response. Additionally, pressure sensors produce varying responses even for the same motion, as differences in execution lead to unique foot pressure patterns - some regions activate more, or the same regions experience different pressure levels. Such variations pose challenges in effectively learning pressure dynamics as these variations do not generalize overall.
To address these challenges, we introduce \emph{Region Importance Learning}, which dynamically adapts to variations using a masking mechanism. Additionally, we incorporate users' biographical data to account for inter-user diversity, enhancing generalization across different individuals.

\subsubsection{Region Importance Learning}\label{sec:channel}
Given a time window of pressure data $\mathbf{x}_{1:W} \in \mathbb{R}^{36 \times W}$, we learn a mask to identify key channels (region) and apply it as a weight to the input pressure data, emphasizing informative regions. \\[-0.7em]

\noindent\textbf{Design of Mask}: We learn this mask via two types of pooling as they are approved to be simple and effective to capture the global temporal context~\cite{pool1,pool2}: 1) global average pooling, $z_{\text{avg}}$ to capture the average pressure response of each channel over time, and 2) global max pooling, $z_{\text{max}}$ to capture the most dominant pressure response in each channel over time as follows:
$
    z_{\text{avg}} = \frac{1}{W} \sum_{t=1}^{W} x_{t}, \;\;\; z_{\text{max}} = \max_{t=1}^{W} x_{t} \;.
$
We then pass both pooled features through a shared multi-layer perceptron (MLP) to obtain $s_{\text{avg}}$ and $s_{\text{max}}$ that learn the importance of pressure sensor channels dynamically. Finally, we fuse them to a learnable importance mask $s$. \\[-0.7em]

\noindent\textbf{Application of Mask}: We apply the learned importance mask $s$ to the input by a simple element-wise dot product over the time dimension, to get the final weighted input $\mathbf{x'}_{1:W} \in \mathbb{R}^{36 \times W}$, and we apply a simple linear layer to transform the weighted input to features $\mathbf{X}_{1:W} \in \mathbb{R}^{D \times W}$ where $D=512$.  \\[-0.7em]

\noindent\textbf{Summary}: While the pooling operations capture global temporal trends, the MLP makes the mask learnable by modeling inter-channel relationships. Eventually, 
the learned importance mask allows {\name} to focus on informative regions, ensuring robustness to variations such as individual foot pressure patterns.

\subsubsection{Integration of User's Biographical Data}\label{sec:bio}
\rev{Proved to be effective~\cite{lsm2, papagei}, we also incorporate \emph{readily available} biographical data (i.e., weight, height, age, shoe size, and gender) from users to improve the model's awareness of individual differences.}   \\[-0.7em]

\noindent\textbf{Usage of Bio Data}: We adopt Feature-wise Linear Modulation (FiLM), a mechanism used in deep learning models to dynamically adjust feature representations based on external inputs, which is widely used in vision and language domains~\cite{film1, film2, film3}.  \\[-0.7em]

\noindent\textbf{Implementation}: Given the output from the region importance learning, $\mathbf{X}_{1:W}$, we perform an affine transformation to the learned pressure feature over time: 
$
    \mathbf{X}^{bio}_{t} = \gamma \odot \mathbf{X}_{t} + \beta, 
$
where $\gamma$ and $\beta$ are feature scaling and shifting parameters that are learned through a two-layer MLP that takes users' biographical data as inputs, and $\odot$ denotes element-wise multiplication.  \\[-0.7em]

\noindent\textbf{Summary}: We leverage FiLM to scale features based on user-specific attributes, as foot pressure responses are dependent on users' physiological information~\cite{gait5}. This step incorporates biographical data into learned features, enhancing the models' awareness of users' differences. 

\subsubsection{Temporal Relationship Learning}\label{sec:temporal}
\rev{To effectively learn the temporal relationships within sequential pressure data, we employ a transformer encoder~\cite{transformer}. Unlike traditional recurrent neural networks, the transformer architecture avoids sequential processing, allowing it to capture dependencies across the entire sequence simultaneously~\cite{bettertrans, papagei, lsm2}. We incorporate position embeddings~\cite{pe} to explicitly encode the order of the time steps, ensuring that the model understands the temporal sequence of pressure events. Moreover, the self-attention mechanism allows the model to dynamically weigh the importance of all input time steps relative to one another. For instance, the model can learn that a pressure pattern from a heel strike is highly relevant to a subsequent pattern during toe-off, enabling it to capture complex, non-local relationships. We empirically utilize a two-layer transformer encoder with four attention heads, and the hidden dimension is set to 512.}

\subsubsection{Muscle Activation Prediction Layer}
After the features are learned, the final layer is another two-layer MLP that takes the features and outputs the final predicted muscle activation levels, $\hat{\mathbf{y}} \in \mathbb{R}^{8 \times W}$.

\subsubsection{Optimization Objectives}\label{sec:loss}
We employ a Mean Squared Error (MSE) loss~\cite{mseloss} that takes both prediction and ground truth muscle activation as follows:
\begin{equation}
    \mathcal{L}^{mse} = \frac{1}{W \cdot M} \sum_{t=1}^{W} \sum_{m=1}^{M} (y_{t,m} - \hat{y}_{t,m})^2, 
\end{equation}
where \( W \) is the time sequence length, \( M \) represents eight muscle groups, \( y_{t,m} \) is the ground truth value at time step \( t \) and channel \( m \), and \( \hat{y}_{t,m} \) is the predicted value. We also introduce a smoothness loss to ensure muscle activation changes gradually over time rather than abrupt, unrealistic fluctuations:
\begin{equation}
    \mathcal{L}^{smooth} = \frac{1}{(W-1) \cdot M} \sum_{t=1}^{W-1} \sum_{m=1}^{M} \left( \Delta \hat{y}_{t,m} - \Delta y_{t,m} \right)^2,
\end{equation}
where \( \Delta y_{t,m} \) and \( \Delta \hat{y}_{t,m} \) represent the changes in ground truth and predicted muscle activation between consecutive time steps, respectively. \\[-0.7em]

\noindent\textbf{Overall}: The objective is a combination of the two losses to jointly optimize for accurate muscle activation predictions and smooth transitions:
$
    \mathcal{L} = \mathcal{L}^{\text{mse}} + \lambda \mathcal{L}^{\text{smooth}},
$
where \( \lambda \) is a hyperparameter that balances between two losses, and we empirically set it to 0.1.

\subsection{Pressure Data Augmentation}\label{sec:aug}
To enhance the diversity of our dataset, we introduce two augmentation techniques for pressure data, (1) \emph{pressure scaling} and (2) \emph{temporal shifting}.
\begin{figure}
\vspace{-0.2in}
	\centering
	\begin{tabular}{c c}
  \hspace{-0.1in}
  
  \subfloat[Pressure Scaling]{\includegraphics[width=0.47\columnwidth,
		scale=1]{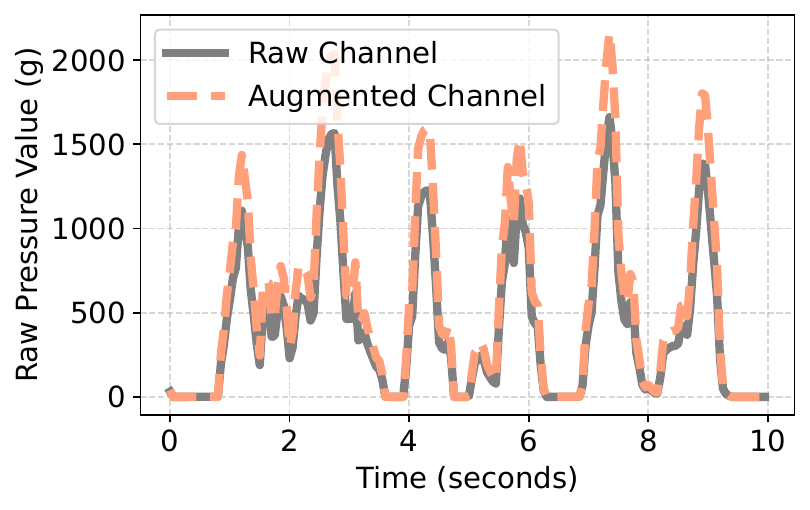}\label{fig:aug_scale}} &
  \subfloat[Temporal Shifting]{\includegraphics[width=0.47\columnwidth,
		scale=1]{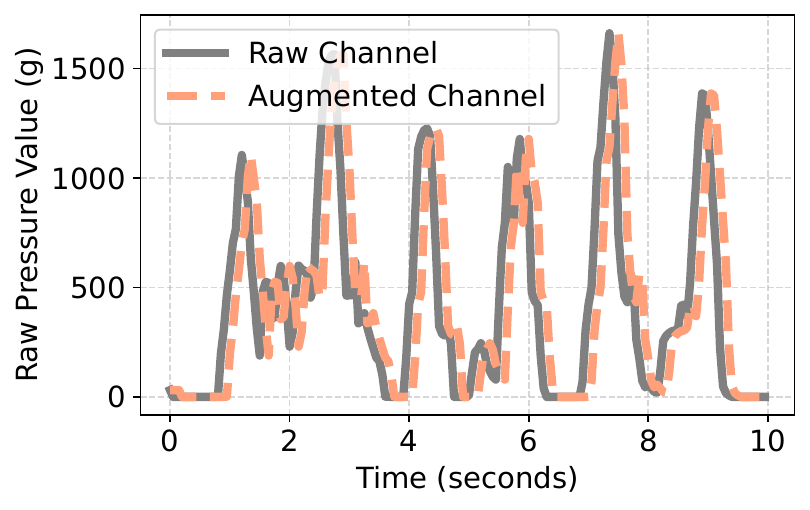}\label{fig:aug_shift}}
	\end{tabular}
    \vspace{-0.1in}
        \caption{Examples of proposed pressure augmentations.}
        \label{fig:aug}
        \vspace{-0.2in}
\end{figure}

\subsubsection{Pressure Scaling} For each pressure sensor channel, the probability of applying a scaling factor $\alpha$ depends on the magnitude of the pressure data. When large magnitude changes are detected  (i.e., $\gt$ 0.3~kg), augmentation is applied with a higher probability, while for near-zero values, the probability is reduced to avoid excessive distortion of sparse data. This ensures that augmentation affects meaningful pressure changes while preserving the natural distribution of low-pressure regions. \\[-0.7em]

\noindent \rev{\textbf{Rationale}: Our technique is motivated by the observed variability in pressure magnitude, which arises from factors like moving styles and maintaining balance, as well as other factors such as shoe types, floor surfaces, and sensor placements. To accurately model these common, real-world conditions without introducing unrealistic data distortions, we empirically selected a scaling factor range of $\alpha \in [0.8, 1.2]$.}

\subsubsection{Temporal Shifting} For each pressure channel, there is a 50\% probability of pressure data shifting forward or backward by a max of $d$ time steps while maintaining the overall sequence structure. We set $p=0.5$ and $d=5$ (250~ms) to ensure that the shifts remain within a reasonable range, preventing excessive delays or advancements while preserving realistic temporal dynamics. We apply this augmentation to all channels.  \\[-0.7em]

\noindent \rev{\textbf{Rationale}: The maximum temporal shift of 250~ms is based on established biomechanical principles~\cite{responsetime} and observations from our user studies. A larger shift could introduce an unrealistic temporal disconnection between pressure data and muscle activation, degrading the model’s performance. The 50\% probability ($p=0.5$) for applying the shift was selected to ensure that the augmentation is not applied to every data point, which helps the model learn both clean and augmented data distributions, thus enhancing its overall robustness.}

\subsubsection{Summary and Benefit}  
\rev{Grounded in empirical observations from our user studies and supported by biomechanical principles, these data augmentation techniques (Fig.~\ref{fig:aug}) introduce controlled variations to enhance the model's ability to generalize. These variations account for key factors such as individual gait differences, dis-coordination, and external conditions, including footwear and sensor placements.}

\section{Experiment Setup}\label{sec:setup}
\begin{figure}
    \centering
\includegraphics[width=0.9\columnwidth, scale=1]{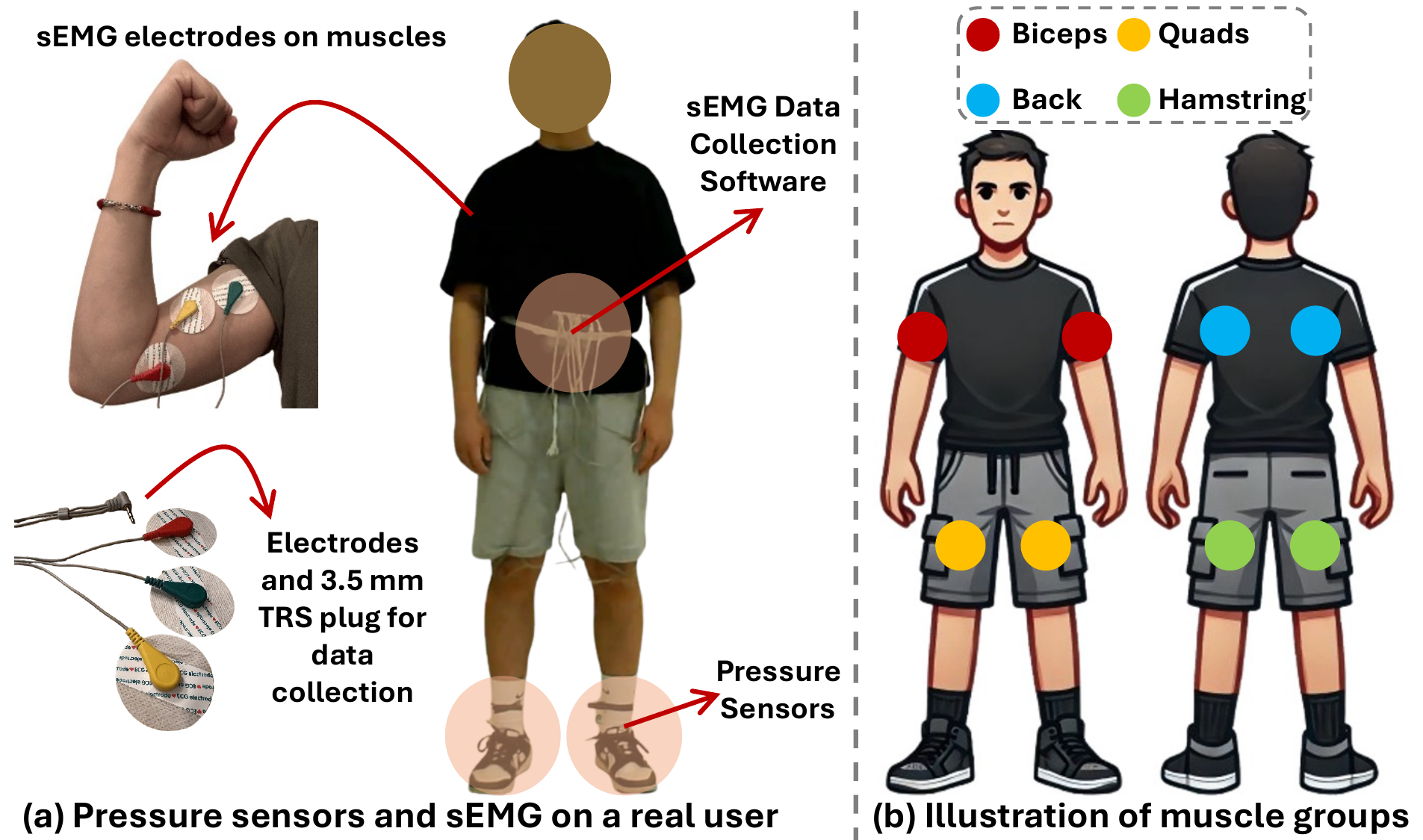}
    \caption{(a) A user wearing both pressure sensors and sEMG electrodes for data collection. The attachments of the electrodes are instructed by professionals while maintaining mobility for user movements. (b) An illustration of muscle groups that are studied in this project.}
    \label{fig:collect}
\end{figure}

\textbf{Data Collection}: \rev{ We recruited 30 subjects (21 males, 9 females), with ages ranging from 22 to 37 years, weights from 39 to 83 kg, and heights from 150 to 186 cm. This cohort also has variations in muscle composition and foot type.} The study was approved by the IRB committee. The ground truth is obtained by surface EMG (sEMG), where electrodes are placed on users' muscle groups (after being wiped with alcohol to increase conductivity) as shown in Fig.~\ref{fig:collect}a. To ensure accurate muscle activation capture while maintaining user mobility, we place the electrodes following guidelines provided by professionals with years of sEMG experience. To ensure the correctness of the ground truth, we first place the electrodes per instructions, and then we ensure the readings are zeros (no activations) when the users relax the target muscle. \rev{Overall, we obtain activations from eight frequently used muscle groups (i.e., left/right biceps, left/right back, left/right quads, and left/right hamstrings as depicted in Fig.~\ref{fig:collect}b). These muscle groups are chosen for coverage of both the upper and lower body and relevance to everyday movements~\cite{mia}. We leave the exploration of monitoring additional muscles for the future (Sec.~\ref{sec:discussion}).}  \\[-0.7em]

\begin{figure}
    \centering
\includegraphics[width=0.9\columnwidth, scale=1]{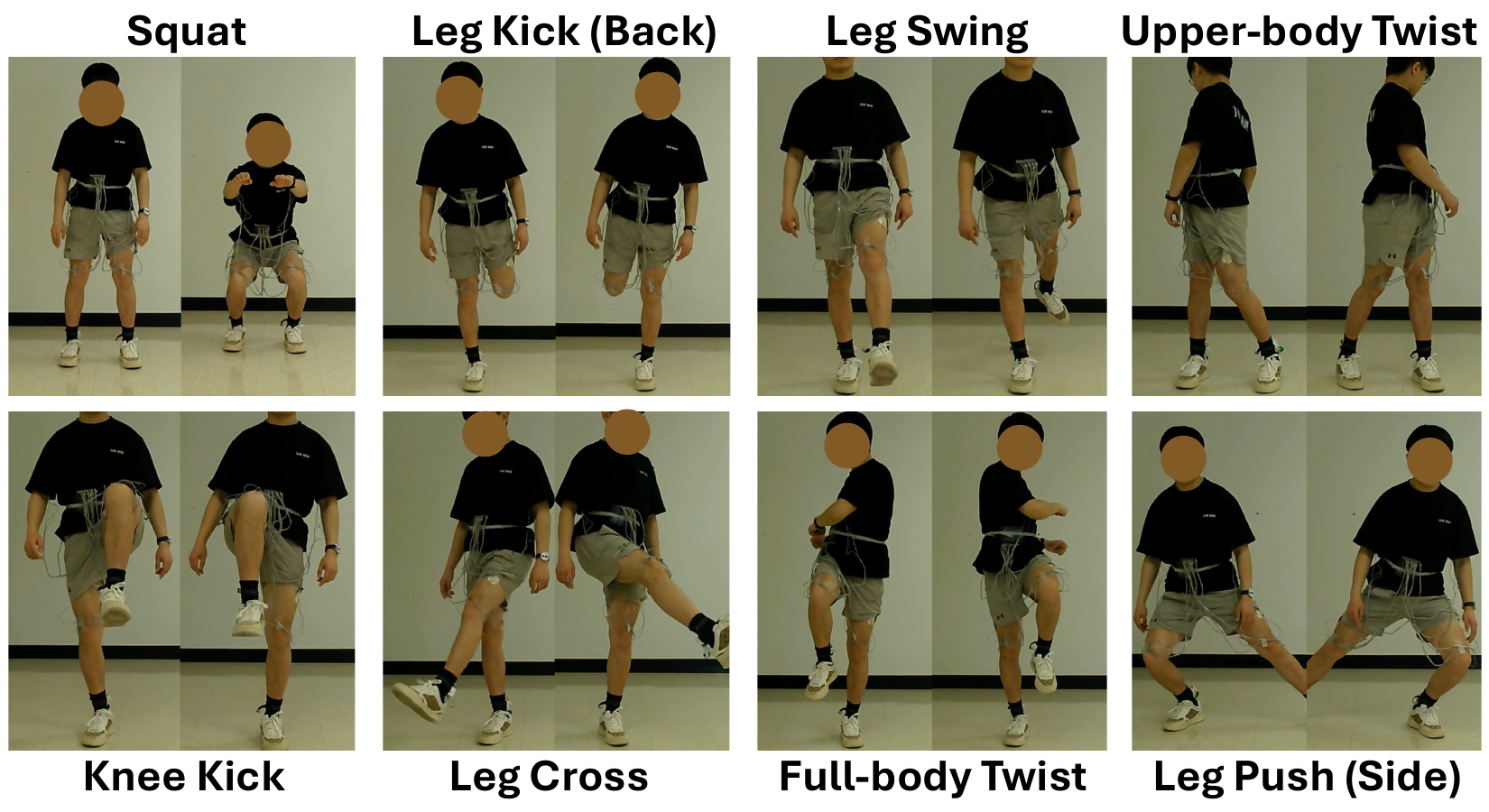}
    \caption{Illustrations of motions (showing eight motions due to space limit). A full list of motions is in Table.~\ref{tab:unseen}.}
    \label{fig:motion}
    \vspace{-0.1in}
\end{figure}

\noindent We carefully select 15 motions (sample motions in Fig.~\ref{fig:motion}) that engage key muscle groups and are commonly seen in real-world fitness and rehabilitation scenarios, following previous works~\cite{mit, mia}. For instance, \emph{squat} activates the quadriceps and hamstrings, essential for lower-body strength in weightlifting and knee/hip rehabilitation. \emph{Upper-body Twist} engages back muscles, vital for rotational power in yoga, and core stability for injury prevention and posture correction. The full list of motions can be found in Table~\ref{tab:unseen}. We ask users to perform each motion for an average of 3 minutes while wearing sEMG electrodes and pressure sensors (Fig.~\ref{fig:collect}a). Each user may take one or more breaks as needed, with each break lasting at least three minutes. During breaks, the pressure sensors are removed and remounted. Total collection time per user is $\approx$2 hours, including setup. \\[-0.7em]

\noindent\textbf{Dataset Details}: We collected approximately 27 hours of pressure data sampled at 20~Hz and corresponding sEMG data at 500~Hz. Synchronized timestamps were used to downsample sEMG data to 20~Hz. Pressure data can reach up to 20~kg per channel, and muscle activation magnitude can reach up to 1000~$\micro V$. To improve model training stability, we convert these values to the range $[0, 1]$, following previous works~\cite{batchnorm, mit, mia}.  \\[-0.7em]

\noindent\textbf{Training Details}: {\name} is implemented across both desktop and smartphones. Our learning modules are built using PyTorch~\cite{pytorch}, with training conducted on a desktop with an Intel i7-8700K CPU, 16GB RAM, and an NVIDIA Quadro RTX 8000 GPU. The input window length $W$ (Sec.~\ref{sec:task}) is set to 20 frames (equivalent to one second). We use the Adam optimizer~\cite{kingma2014adam} with a batch size of 512 and a learning rate of 1e-4. To mitigate over-fitting, we apply L2 regularization~\cite{l2reg} with a coefficient of 0.01. The dataset is augmented twice using the proposed scaling and shifting techniques for training. We opt for these parameters by grid searches. We also deploy the model for inference on Samsung S21 Ultra and Google Pixel 7 Pro using PyTorch Lite~\cite{pylite}.  \\[-0.7em]

\noindent\textbf{Data Split}: \rev{Unless otherwise noted, we adopt a leave-one-user-out (inter-user) validation protocol. The model is trained on data from 29 users and evaluated on the held-out user. This strategy assesses {\name}'s ability to generalize to unseen individuals, demonstrating robustness and adaptability in real-world scenarios.}  \\[-0.7em]

\noindent\textbf{Metrics}: Following previous works~\cite{chen2024exploring, mia, mit}, we use Root Mean Squared Error (RMSE) as the primary evaluation metric to assess the accuracy of muscle activation predictions. RMSE reports the differences between predicted and ground truth values. To calculate RMSE for one muscle group, we employ the equation: $\sqrt{\frac{1}{T} \sum_{t=1}^{T} \left( y(t) - \hat{y}(t) \right)^2}$, where $T$ denotes the length, $y(t)$ and $\hat{y}(t)$ denote ground truth and predicted muscle activation, respectively. We also utilize Pearson correlation~($\rho$) in certain cases to show the correlation between ground truth and predicted values.
The final metrics are reported by averaging across eight muscle groups.

\begin{figure*}
    \centering
    \begin{minipage}{0.59\textwidth}
        \centering
        \includegraphics[width=\textwidth]{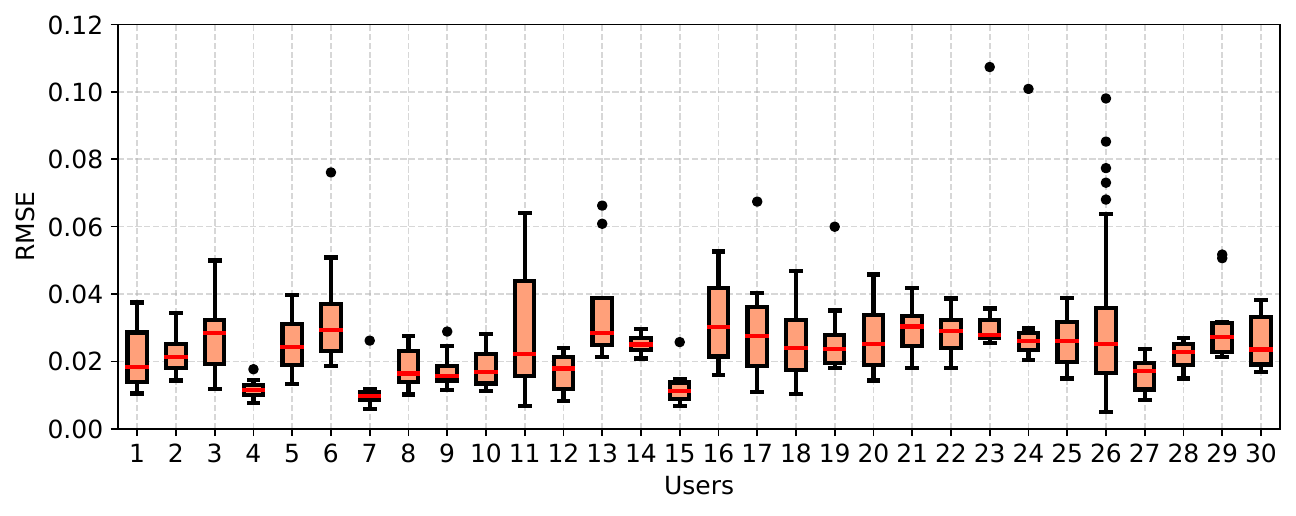}
        \vspace{-0.35in}
        \caption{\rev{{\name} is stable across 30 unseen users.}}
        \label{fig:overall}
    \end{minipage}
    \hfill
    \begin{minipage}{0.4\textwidth}
        \centering
        \includegraphics[width=\textwidth]{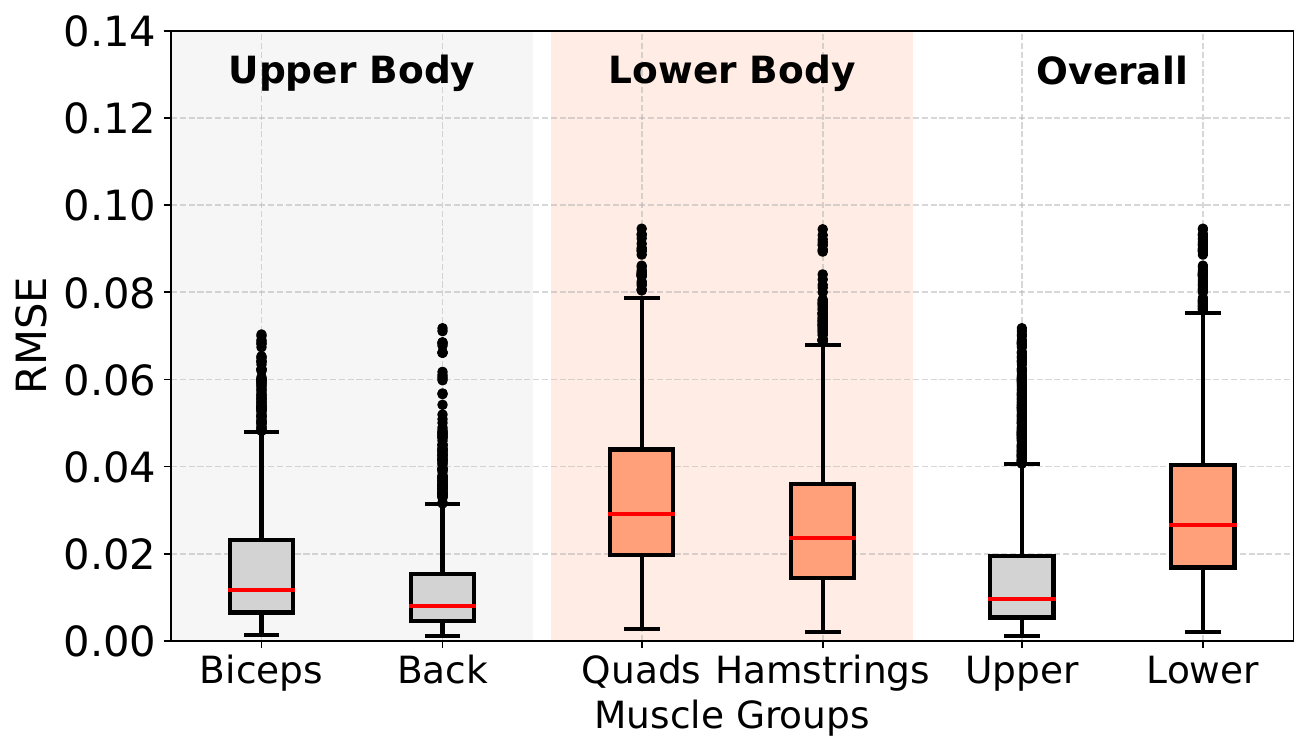}
        \vspace{-0.3in}
        \caption{Performance over muscle groups.}
        \label{fig:musclegroups_rmse}
    \end{minipage}
\end{figure*}

\section{Performance Evaluation}\label{sec:eval}
\rev{We evaluate {\name}'s generalization to users and motions, as well as its robustness across various conditions, including different ground surfaces, and user demographics. We also experimentally validate our theoretical foundation by restricting visible motion, and the effectiveness of our proposed techniques by comparing {\name} against state-of-the-art methods. We conduct an analysis of the model's complexity, power consumption, and inference latency on smartphones, along with case studies including muscle imbalance detection and an in-situ study, demonstrating its applicability.}

\begin{figure*}[htb]
    \centering
    \begin{minipage}{0.51\textwidth}
        \centering
        \begin{tabular}{c}
    \hspace{-0.15in}
  \includegraphics[width=1\columnwidth,
		scale=1]{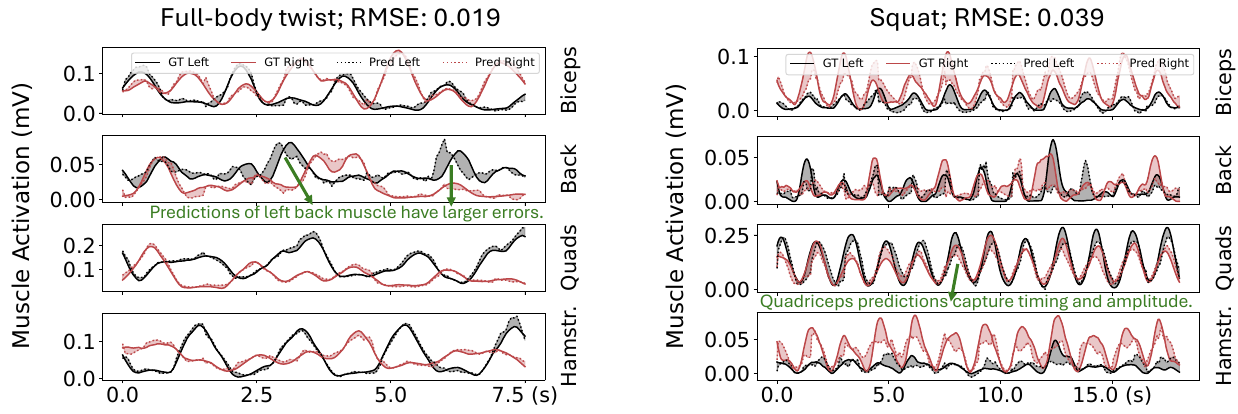}
	\end{tabular}
    \vspace{-0.1in}
        \caption{\hao{Qualitative results.} \setlength\fboxsep{0.8pt}\colorbox{Gray}{\strut gray} and \setlength\fboxsep{0.8pt}\colorbox{red!20}{\strut red} areas indicate errors from left- and right-side muscles, respectively.}
        \label{fig:qual}
    \end{minipage}
    \hfill
    \begin{minipage}{0.45\textwidth}
    \vspace{-0.1in}
    \centering
	\begin{tabular}{c c}
    \hspace{-0.2in}
  \subfloat[Shoe Type]{\includegraphics[width=0.64\textwidth,
		scale=1]{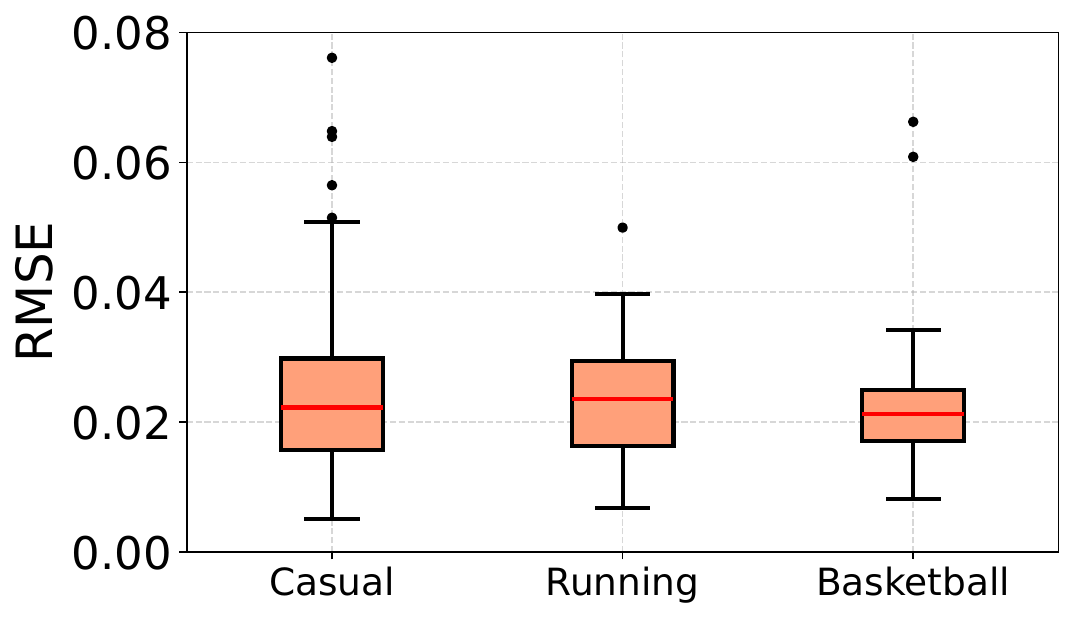}\label{fig:shoetype}} &
        \hspace{-0.1in}
  \subfloat[Surface Type]{\includegraphics[width=0.367\textwidth,
		scale=1]{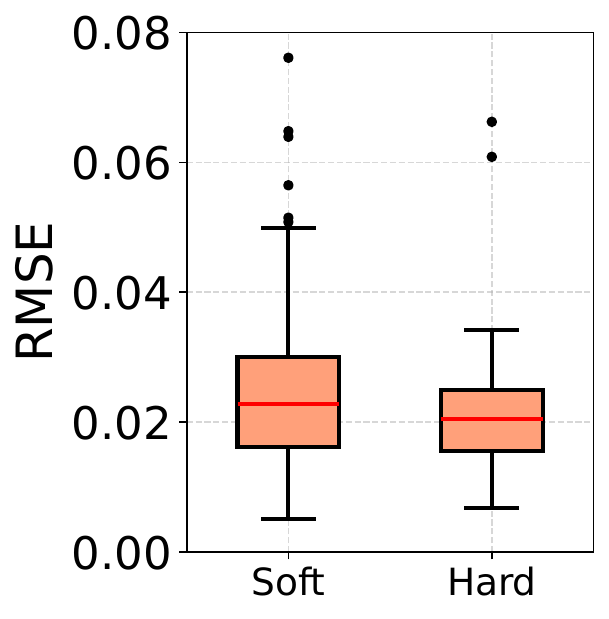}\label{fig:surfacetype}}
	\end{tabular}
    \vspace{-0.1in}
        \caption{{\name} is robust to operating surfaces.}
    \end{minipage}
\end{figure*}

\subsection{Overall Performance}
\noindent\textbf{Unseen Users}: Fig.~\ref{fig:overall} depicts {\name}'s performance across 30 unseen users. Evidently, {\name} achieves decent performance across all users, with a mean RMSE of 0.025. We note that a few users (e.g., User 9) exhibit slightly higher errors than others due to unintentional body motions, such as scratching their heads or hair, during the data collection. Such incidental movements can be filtered out by adding a smartwatch~\cite{liu2023facetouch,yang2024maf} to detect these motions, improving the robustness. In a nutshell, these results are from unseen users, meaning the testing user was not included in the training data. This ensures that the evaluation reflects {\name}'s ability to generalize to new individuals rather than memorizing specific user patterns, thanks to the proposed techniques for effective feature representations.  \\[-0.7em]

\noindent\textbf{Muscles Groups}: Fig.\ref{fig:musclegroups_rmse} depicts the decoupled performance from lower and upper body muscles. Interestingly, errors are higher for lower-body muscles. When users perform certain motions (e.g., leg swing), upper-body muscles are minimally activated, leading to low RMSE errors since most activation values are close to zero. This aligns with the observations from previous work~\cite{mit}. Nevertheless, we believe {\name} exhibits stable performance across the entire body, validating the connection between foot dynamics and muscle activation. \\[-0.7em]

\noindent\textbf{Qualitative Results}: Fig.~\ref{fig:qual} presents two qualitative examples with a low (0.019) and relatively high RMSE (0.039) from full-body twist and squat, respectively. Evidently, the predicted activation across all muscle groups closely aligns with the ground truth.

\subsection{Robustness Study}\label{sec:robust}
\noindent\textbf{Impact of Operating Surfaces}: Our sensor platform utilizes flexible film pressure sensors (Sec.~\ref{sec:platform}) that require a relatively hard surface to register pressure responses effectively. To evaluate the robustness of {\name}, we asked users to wear different shoe types (casual, running, and basketball, each with varying insole materials) and perform motions on both hard surfaces (concrete, marble) and a relatively softer surface (carpet). As shown in Fig.\ref{fig:shoetype} and Fig.\ref{fig:surfacetype}, while the softer surfaces absorbed some forces and influenced sensor responses, {\name} maintained stable overall performance across different shoe types and surface conditions. We believe such stability of {\name} comes from the proposed data augmentation, which amplifies or attenuates pressure responses, ensuring reliable performance despite variations in footwear and ground surfaces. \\[-0.7em]

\begin{figure}
	\centering
{\includegraphics[width=0.76\columnwidth,
		scale=1]{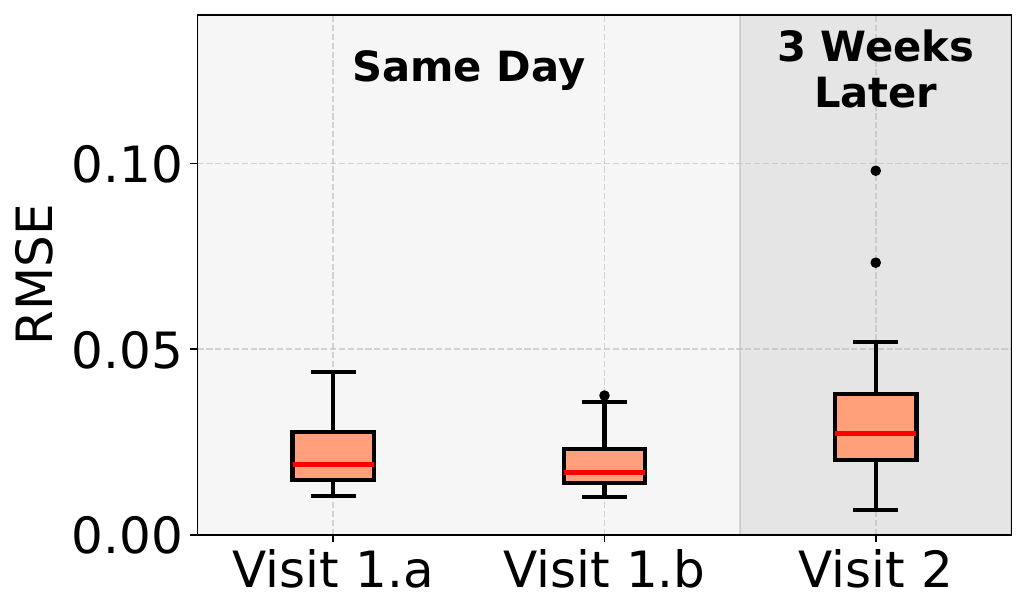}}
        \vspace{-0.1in}
        \caption{{\name} demonstrates robustness and stability over sensor placements (Visit 1.a and 1.b) and time.}
        \label{fig:remountandlong}
\end{figure}

\begin{figure}[htb]
	\centering
{\includegraphics[width=0.97\columnwidth,
		scale=1]{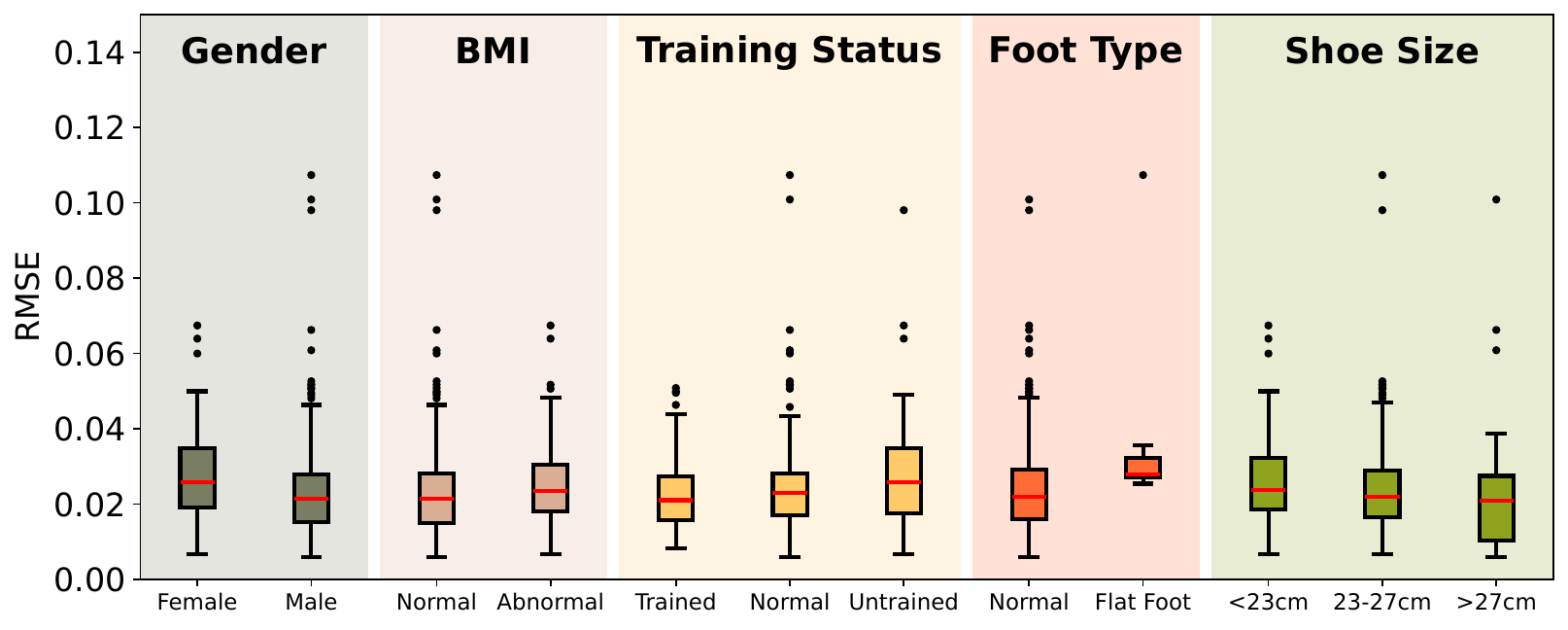}}
        \vspace{-0.2in}
        \caption{\rev{Performance by demographics.}}
        \label{fig:demographics}
\end{figure}

\begin{figure}
	\centering
{\includegraphics[width=0.97\columnwidth,
		scale=1]{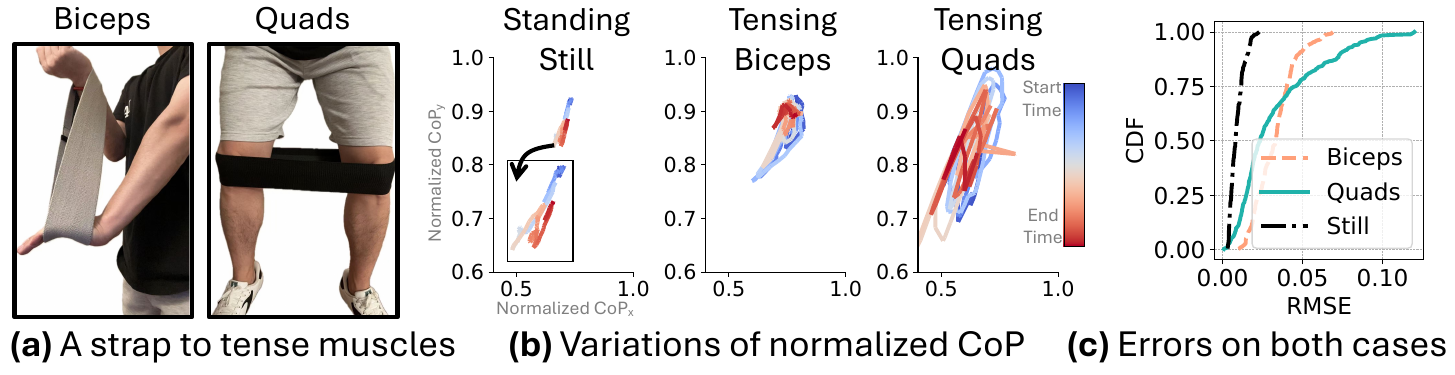}}
        \vspace{-0.1in}
        \caption{Relationship of tensing muscles only (no movements) and CoP from pressure insoles.}
        \label{fig:copfinal}
\end{figure}

\noindent\textbf{Impact of Sensor Locations}: Since people repeatedly wear shoes every day, 
it is essential to account for the impact of potential sensor placement shifts over time. We utilized the breaks users took during the study, during which the pressure sensors were removed and remounted. This process naturally introduced slight variations in sensor placement within the shoes, allowing us to assess how these changes affect performance. As reported in Fig.~\ref{fig:remountandlong}, we do not see big variations in errors before (Visit 1.a) and after (Visit 1.b) the break. We believe the reasons are two-fold: 1) sensors are snugly fit in shoes; 2) in the cases of shifted sensor placements (different regions got activated), our proposed region importance learning focuses on informative regions, making {\name} robust to such variations. \\[-0.7em]

\noindent\textbf{Stability Over Time}: To examine the potential impact of time-related factors (e.g., weight fluctuations), we invited the same users from the \emph{Impact of Sensor Locations} study to return for another session after three weeks. While it is a relatively short period with minimal expected physiological changes, it is encouraging to see stable performance over time (Fig.~\ref{fig:remountandlong}). This suggests that {\name} does not merely memorize patterns from the initial session, especially since the user is unlikely to perform identical motions across both visits.  \\[-0.7em]

\noindent\rev{\textbf{Impact of Demographics}: To examine the impact of gender, Body Mass Index (BMI), muscle composition, and shoe size, we categorize users different groups: (1) male and female users, (2) normal BMI (i.e., $[18.5, 25]$, based on standard criteria~\cite{bmi}) and abnormal BMI (outside this range), (3) muscle composition based on self-reports, where users are divided into trained (7 of subjects reported frequent exercise and greater muscle mass), untrained individuals (5 of subjects reported little to no exercise), and normal individuals (reported engaging in light to moderate activities), and (4) normal foot type and flat foot. (5) Finally, we categorize our users into three groups based on their foot length: ``<23cm'' (N=7), ``23cm-27cm'' (N=16), and ``>27cm'' (N=7). Fig.~\ref{fig:demographics} demonstrates stable performance across these factors, highlighting {\name}’s effectiveness for diverse demographics and physical profiles.}  \\[-0.7em]

\vspace{-0.1in}

\subsection{Muscle-to-Pressure Validation}\label{sec:m2p_val} \rev{While pressure variations encode motion features, we now study whether plantar pressure encodes muscle activation without visible motions: \emph{Users tense their biceps or quadriceps using a stretch band (Fig.~\ref{fig:copfinal}a) while remaining as still as possible}. Compared with relaxed standing, these no-motion activations produced variations in center-of-pressure (CoP), likely reflecting involuntary micro-adjustments required to maintain balance (Fig.~\ref{fig:copfinal}b). As depicted in Fig.~\ref{fig:copfinal}c, they were still predictable relative to the ground-truth sEMG. 
We believe the results suggest that insole pressure signals are sensitive to capture such subtle physiological effects via the link of CoP (Sec.~\ref{sec:just}). This finding extends the applicability of {\name} beyond overt movements, supporting the potential of pressure insoles as a practical tool for monitoring a broader range of muscle activities.} 
\hao{We clarify that our assumption excludes only substantial external loads that fundamentally alter whole-body dynamics and redistribute weight in ways that obscure the intrinsic relationship between muscle activation and plantar pressure. Minor external tools, such as light resistance bands, remain within the scope of our study. In particular, the resistance band used in this study introduces only small, elastic tension and does not impose meaningful external loading on the lower body or induce observable shifts in CoM or CoP based on our measurements. As a result, we believe such tools do not violate the underlying biomechanical assumptions introduced in Sec.~\ref{sec:just}. In contrast, larger external loads (e.g., a 5~kg dumbbell) that significantly redistribute body weight or constrain posture fall outside our current framework. Addressing these cases, for example, by explicitly modeling external forces or incorporating additional sensing modalities, represents an important direction for future work to further broaden the applicability of {\name}. }

\vspace{-0.2in}

\subsection{Performance of Unseen Motions}\label{sec:unseenmotion}
We design a leave-one-motion-out protocol to validate {\name}'s generalization on unseen motions, where 14 motions are used for training, and the remaining motion is only for testing. Table~\ref{tab:unseen} reports the performance. {\name} achieves an average RMSE of 0.027 and an average correlation value of 0.504 across unseen motions. 
Furthermore, we roughly categorize these motions as full-body, lower-body, and upper-body motions based on the major muscle groups used in the motions. Evidently, {\name} performs stably over these motions. 
In particular, {\name} performs well on upper-body motions, suggesting our biomechanical link (Sec.~\ref{sec:just}) is practically valid for the upper body muscles. 
Overall, while these motions are not present in the training data, {\name} successfully generalizes to them, suggesting that the model learns a meaningful relationship between muscle activation and foot pressure dynamics rather than memorizing specific motion patterns. \rev{We also validate with in-situ activities in Sec.~\ref{sec:case}.}

\begin{table}[htb]
    \centering
\resizebox{0.8\columnwidth}{!}
{
\begin{tabular}{ l c c} 

 \multirow{2}{*}{\textbf{Motion}} & \multicolumn{2}{c}{ \textbf{{\name}}} \\
\cmidrule(lr){2-3}
  & \textbf{RMSE} $\downarrow$ & $\rho$ $\uparrow$  \\
 
\cmidrule(lr){1-1}
\cmidrule(lr){2-3}
\rowcolor{lightred}
Knee Kick & 0.029 & 0.515 \\[0.17cm]
\rowcolor{lightred}
Leg Cross & 0.023 & 0.475 \\[0.17cm]
\rowcolor{lightred}
Leg Kick (B) & 0.027 & 0.580 \\[0.17cm]
\rowcolor{lightred}
Leg Swing (F/B) & 0.025 & 0.507 \\[0.17cm]
\rowcolor{lightred}
Leg Swing (S) & 0.022 & 0.570 \\[0.17cm]
\rowcolor{lightred}
Leg Push (F) & 0.025 & 0.589 \\[0.17cm]
\rowcolor{lightred}
Leg Push (S) & 0.029 & 0.659 \\[0.17cm]
\rowcolor{lightgreen}
Arm Swing & 0.018 & 0.526 \\[0.17cm]
\rowcolor{lightgreen}
Open Arm \& Chest Expansion & 0.018 & 0.550 \\[0.17cm]
\rowcolor{lightgreen}
Swing a Tennis Racket & 0.037 & 0.381 \\[0.17cm]
\rowcolor{lightgreen}
Upper-body Twist & 0.033 & 0.438 \\[0.17cm]
\rowcolor{lightblue}
Squat & 0.033 & 0.488 \\[0.17cm]
\rowcolor{lightblue}
Strech (S) & 0.033 & 0.445 \\[0.17cm]
\rowcolor{lightblue}
Full-body Twist & 0.036 & 0.335 \\[0.17cm]
\rowcolor{lightblue}
Jack Jump & 0.023 & 0.504 \\[0.17cm]
Average & 0.027 &  0.504 \\
\bottomrule
\end{tabular}
}
\caption{Performance on unseen motions. 
F, B, and S denote forward, backward, and side, respectively. $\downarrow$ indicates lower is better and $\uparrow$ indicates higher is better. Shaded areas separate motions into \colorbox{lightred}{lower-body motions}, \colorbox{lightgreen}{upper-body motions}, and \colorbox{lightblue}{full-body motions}.}
    \label{tab:unseen}
    \vspace{-0.2in}
\end{table}

\subsection{Ablation Study}\label{sec:ablation}

\begin{table}
    \centering

    \resizebox{0.9\columnwidth}{!}
    {
    \begin{tabular}{l S[table-format=-2.2] S[table-format=-2.2]} 
    \textbf{Setting} & \textbf{$\Delta$RMSE (\%)}  & \textbf{$\Delta\rho$ (\%)} \\

    \cmidrule(lr){1-1}
    \cmidrule(lr){2-3}
    \rowcolor{Gray}
    \textit{{\name}} & 0.00 & 0.00 \\[0.17cm]
    \;  $\setminus$ Region Importance Learning & -8.65 &-11.08 \\[0.17cm]
    \;  $\setminus$ Bio-FiLM & -5.93 &-8.47 \\[0.17cm]
    \;  $\setminus$ Scale Augmentation & -4.68 &-7.18 \\[0.17cm]
    \;  $\setminus$ Shift Augmentation & -4.69 &-5.03 \\[0.17cm]
    \;  $\setminus$ Any Augmentation & -5.87 &-10.84 \\[0.17cm]
    \;  $\setminus$ Smooth loss  & -3.85 &-2.30 \\
    \bottomrule
    \end{tabular}
    }
    \caption{Ablation study. $\setminus$ denotes a model variant without a certain component. We directly report the performance relative to {\name} for all the metrics. Negative numbers suggest worse performance.}
    \vspace{-0.25in}
    \label{tab:ablation}
\end{table}

\noindent An ablation study (Table~\ref{tab:ablation}) confirmed the contribution of each component. Removing \emph{Region Importance Learning} was most detrimental, increasing RMSE by 8.65\% and reducing correlation by 11.08\%, which highlights its critical role in feature extraction. Excluding biographical data, data augmentation, or smooth loss also degraded performance by different margins, reinforcing their values for improving {\name}'s robustness.

\subsection{Comparisons with Existing Methods}
\begin{table}
    \centering
\resizebox{0.77\columnwidth}{!}
{
\begin{tabular}{ l cc} 

 \textbf{System}  & \textbf{RMSE} $\downarrow$ & $\rho$ $\uparrow$ \\
\cmidrule(lr){1-1}
\cmidrule(lr){2-3}
Muscle in Action~\cite{mia} & 0.035  & 0.473\\[0.17cm]
Muscle in Time~\cite{mit} & 0.031  & 0.501\\[0.17cm]
DANN~\cite{ganin2016domain} & 0.029  & 0.512\\[0.17cm]
\rowcolor{Gray}
{\name} & 0.025 & 0.589\\
\bottomrule
\end{tabular}

}
\caption{Comparisons with existing methods.} %
    \label{tab:compare}
    \vspace{-0.2in}
\end{table}

\noindent \rev{We compare {\name} with two methods. While \emph{Muscle In Action}~\cite{mia} utilizes raw videos to predict muscle activation, \emph{Muscle in Time}~\cite{mit} utilizes human meshes, which eliminates the potential noise in raw videos. We also compared against Domain-Adversarial Neural Networks (DANN)~\cite{ganin2016domain} that learn invariant features from domains (users). Table~\ref{tab:compare} shows {\name} performs better than these counterparts.}

\subsection{Complexity, Power, and Latency}\label{sec:power}
\noindent\rev{\textbf{Complexity}: The {\name} model has 7.9M parameters ($\approx$32 MB, FP32) with a computational cost of $\approx$300M FLOPs (comparable to mobile-friendly architectures such as MobileNet~\cite{mobilenets}). } \\[-0.7em]

\noindent\textbf{Power}: We implement {\name} via PyTorch Lite~\cite{pylite} on Samsung S21 Ultra and Google Pixel 7 Pro. To quantify power usage, we run the trained model continuously on pressure data for two hours, recording the remaining battery percentage every 30 minutes. We use the discharge rate as a proxy for power consumption. Samsung S21 Ultra and Pixel 7 Pro discharged at 10\% and 12\% per hour, respectively. The power consumption of the hardware platform is discussed separately in Sec.~\ref{sec:platform}.  \\[-0.7em]

\noindent\textbf{Latency}: {\name} processes a time window of 20 pressure data, incurring a one-second delay (sampling at 20 Hz) only at the start. To meet real-time requirements, we optimize the data loading pipeline to reuse previous data, ensuring instantaneous processing at any given time. With this modification, the average model inference latency is 33 ms on the Samsung S21 Ultra and 25 ms on the Google Pixel 7 Pro.

\subsection{Case Study}\label{sec:case}

\begin{figure}[t]
    \centering
    \vspace{-0.15in}
    \begin{tabular}{c c}
\subfloat{\includegraphics[width=0.45\columnwidth]{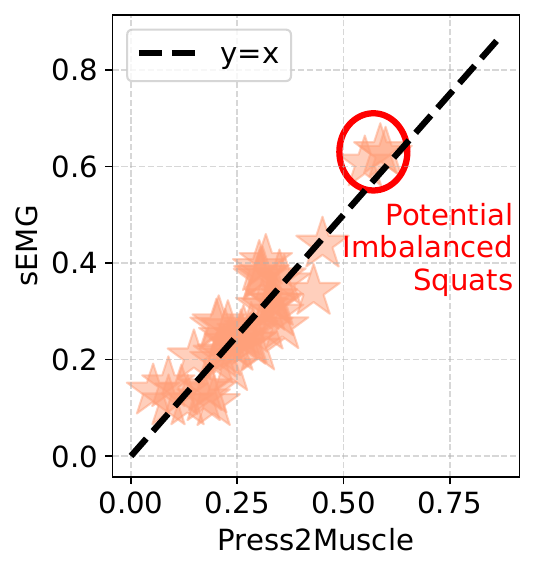}\label{fig:casestudy}} &
\subfloat{\includegraphics[width=0.45\columnwidth]{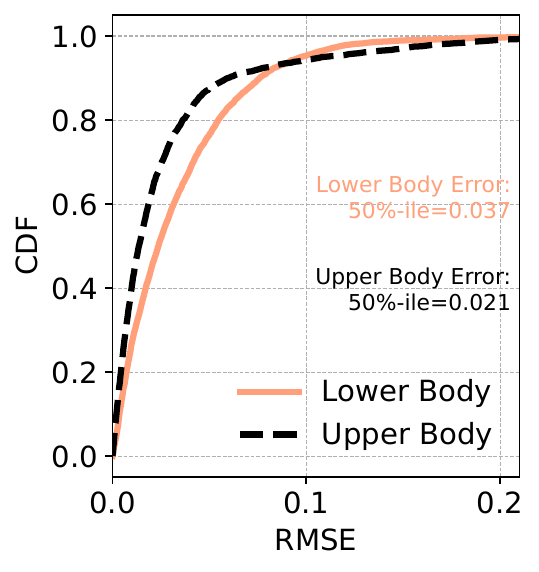}\label{fig:freeliving}}
    \end{tabular}
    \vspace{-0.2in}
    \caption{(a) Muscle Imbalance. (b) \hao{Free-living}.}
\end{figure}

\noindent\textbf{Muscle Imbalance:} We showcase {\name} in detecting muscle imbalance using \emph{Squat}, a motion ideally requiring equal left-right muscle contribution (Fig.~\ref{fig:motion}). We compute an imbalance score by measuring the absolute activation difference between left and right muscles, normalizing by total activation, and averaging across time and paired muscle groups. This results in a score ranging from 0 (fully balanced) to 1 (fully imbalanced). Fig.~\ref{fig:casestudy} shows a 0.92 correlation between the scores calculated from sEMG and {\name}-based muscle activations from 45-minute squatting recordings. \rev{An RMSE of 2.5\% corresponds to a 0.025 error on the 0–1 activation scale. Given that left–right asymmetries of 5–10\% (0.025–0.05 mV at 0.5 mV activation) are considered indicators of injury risk~\cite{imp1, imp2}, {\name} achieves a finer resolution than these thresholds, accurately identifying squats with an imbalance score above 0.5. We believe {\name} holds significant potential for clinical integration, enabling injury prevention and offering deeper insights. We leave this direction for future work.} \\[-0.7em]

\noindent\textbf{Free-living:} To explore the real-world applicability of {\name} during daily activity, we conducted a 30-minute session in which subjects performed a variety of everyday tasks (e.g., walking, reading a book, typing, climbing and descending stairs, taking phone calls, stretching the body, throwing a ball, and retrieving objects). Ground truth sEMG data were collected throughout the session, and segments involving external loads (e.g., carrying heavy objects) were excluded from analysis. \hao{As depicted in Fig.~\ref{fig:freeliving}, {\name} captures muscle activations well, with median errors of 0.037 (lower body) and 0.021 (upper body). Since the free-living evaluation involves diverse motions performed in random order, a higher error is expected. Despite this increased variability, the observed performance remains reasonable and highlights {\name}'s potential for health monitoring and fitness tracking in daily life.}

\section{Discussion and Future Work}
\label{sec:discussion}

\noindent\hao{\textbf{Positioning and Scope of {\name}.}
As mobile health technologies gain widespread adoption~\cite{health_blood3, health_blood1, health_res4, health_stress1}, continuous and unobtrusive monitoring of human physiology has become increasingly important. 
While muscle activation plays a central role in movement quality, injury prevention, and rehabilitation, it remains underexplored in everyday settings due to the practical limitations of existing sensing modalities. {\name} is the first system to demonstrate that muscle activation can be inferred from foot pressure patterns captured by pressure insoles, offering a novel and highly unobtrusive alternative to conventional sensing approaches such as sEMG, prioritizing long-term comfort, wearability, and deployment feasibility in daily-life scenarios.}  \\[-0.7em]

\noindent\hao{\textbf{Limitations and Applicability.}
Our theoretical framework links muscle activation to plantar pressure changes through whole-body biomechanical coupling during non-tool-assisted motions.
While this assumption holds for a wide range of standing and ambulatory activities, we acknowledge that upper-body muscle activations may not always produce measurable changes in CoM or CoP, particularly in scenarios such as seated upper-body movements or highly constrained postures.
In such cases, the biomechanical pathway from muscle activation to foot pressure becomes attenuated, limiting the applicability of insole-based sensing alone.
Extending the framework to tool-assisted motions or seated activities, therefore, remains an important direction for future work.} \\[-0.7em]

\noindent\hao{\textbf{Future Directions.}
Building on these limitations, we plan to explore several extensions to broaden the applicability of {\name}. First, we will investigate hybrid sensing strategies that combine pressure insoles with lightweight wearable sensors to better capture upper-body activations in seated or low-mobility scenarios. Second, while our model performed well across the subjects with diverse demographics, we will continue to expand our scope to include more diverse subjects. Finally, we envision adapting {\name} to rehabilitation-oriented settings, particularly for elderly users, where seated exercises and controlled upper-body movements are common. In these contexts, pressure insoles may serve as a complementary signal within a multimodal sensing framework rather than a standalone solution.}  \\[-0.7em]

\noindent\hao{\textbf{Usability and User Experience.}
To assess the practical trade-offs between {\name} and sEMG, we conducted a usability study with 15 participants. On a 10-point Likert scale, {\name} achieved higher scores in wearability and ease of use, as summarized in Box~\ref{box:user_experience}. Representative feedback is also included in the same box. These findings highlight a fundamental trade-off: while sEMG provides precise muscle measurement in clinical environments, {\name} provides a more practical and user-friendly solution for continuous, long-term monitoring in free-living environments, aligning well with applications such as daily activity feedback, rehabilitation monitoring, and fitness coaching.}

\vspace{0.2in}

\refstepcounter{boxctr}
\label{box:user_experience}
\noindent\hao{\textbf{Box~\theboxctr: User Experience Comparison and Feedback.}}

\begin{tcolorbox}[
    colback=white,
    colframe=black,
    arc=6pt,
    boxrule=0.8pt,
    left=3mm,
    right=3mm,
    top=2mm,
    bottom=2mm
]
\small
\begin{center}
\begin{tabular}{lcc}
\textbf{Aspect} & \textbf{{\name}} & \textbf{sEMG} \\
\midrule
Wearability (1--10) & 9.2 & 2.5 \\
Ease of Use (1--10) & 9.5 & 3.1 \\
Long-Term Comfort & High & Low \\
Deployment Feasibility & High & Limited \\
Reported Skin Irritation & None & Occasional \\
\bottomrule
\end{tabular}
\end{center}

\hao{\textbf{Wearability:}
``The insoles felt natural during the activities, and after a while, I hardly noticed them. But the sEMG electrodes were more difficult to attach and became uncomfortable over time.''}

\vspace{0.3em}
\hao{\textbf{Ease of Use:}
``Once they were in my shoes, everything felt like normal footwear. By comparison, setting up the sEMG sensors required much effort, and I would be reluctant to use them beyond the study setting.''}

\vspace{0.3em}
\hao{\textbf{Willingness to Integrate:}
``I could imagine using the insoles regularly without changing my habits, and I would like to see how they monitor my daily activities and provide feedback over time.''}
\end{tcolorbox}

\vspace{-0.1in}
\section{Related Work}\label{sec:related}

\noindent\textbf{Pressure-related Sensing}: Pressure has been creatively explored in various applications. Prior work has demonstrated its effectiveness in pose reconstruction~\cite{PressInPose, soleposer, pressure_motion1,wu2024dual,pressure_pose1, pressure_pose2}, authentication~\cite{pressure_auth1, pressure_auth2}, and human-computer interaction~\cite{pressure_interaction1, pressure_interaction2}. PressInPose~\cite{PressInPose} employs insole pressure sensors and inertial sensors to reconstruct full-body human pose, utilizing the relationship between foot pressure distribution and body posture. 
PressureNet~\cite{pressure_pose1} leverages pressure-sensitive mattresses to estimate human poses, and CAvatar~\cite{pressure_pose2} employs pressure carpets to reconstruct body posture. 
TouchPass~\cite{pressure_auth1} introduces a pressure-based authentication system, analyzing unique vibrations generated by finger pressure on touchscreens for identity verification. These applications demonstrate the versatility of pressure, but they primarily focus on external movement analysis. 
In contrast, {\name} introduces a novel application of muscle activation sensing with insole pressure sensors. \\[-0.7em]

\noindent\textbf{Muscle Activation Sensing}: sEMG~\cite{emg1} provides highly accurate muscle activation monitoring but is costly, may cause skin irritation, and is impractical for continuous daily use~\cite{emg2}. Video-based approaches~\cite{mia, mit} estimate muscle activation by tracking external body movements, yet videos are constrained by environmental conditions, prone to occlusions, and pose privacy concerns, limiting the practicality. In contrast, {\name} explores muscle activation sensing using insole pressure sensors, leveraging the natural biomechanical link between foot pressure dynamics and muscle recruitment, offering a more accessible and unobtrusive alternative.  \\[-0.7em]

\noindent\textbf{Physiological Sensing}: Mobile health has gained significant popularity, enabling the continuous monitoring of various physiological signals through wearable devices and wireless signals. Most existing work in this domain has focused on tracking physiological parameters, such as respiration~\cite{chen2021movi, zheng2022catch, health_res1, health_res2, health_res3, health_res4}, cardiac functions~\cite{headfi, zhang2022can, health_blood1, health_blood2,health_bp2, health_par,zhang2022wifi, zhou2025know,chan2019contactless,pai2021hrvcam}, and affective sensing~\cite{health_sleep1, health_stress1, health_emotion1,liu2015tracking}. MSense~\cite{health_res4} introduces a robust respiration estimation system that remains accurate under motion interference. EQ-Radio~\cite{health_emotion1} showcases emotion recognition using RF signals, eliminating the need for direct physical contact. eBP~\cite{health_blood1} leverages in-ear Photoplethysmography to provide a comfortable method for blood pressure estimation. While these works have undoubtedly pushed the boundaries of mobile health, {\name} is the first to focus on important yet overlooked muscle activation signals.

\vspace{-0.1in}
\section{Conclusion}\label{sec:conclusion}

\rev{We introduced {\name}, a novel pressure insole system for unobtrusive, full-body muscle activation sensing. Validated on 30 users, our system achieved an RMSE of 0.025, demonstrating that foot pressure is an effective proxy for muscle activation and offering a scalable wearable solution for mobile health and fitness.}

\section*{Acknowledgements}
Our gratitude goes to the anonymous reviewers and shepherds for their invaluable feedback. This research has been partly funded by NSF grants: CAREER-2046972.

\bibliographystyle{ACM-Reference-Format}
\bibliography{ref}

\end{document}